\documentclass[useAMS,usenatbib]{mn2e}

\usepackage{color}
\usepackage{lscape,graphicx}
\usepackage{amsmath}
\usepackage{journal_shortcuts}
\usepackage{amssymb}
\usepackage{multirow}
\usepackage{afterpage}
\usepackage{ulem}
\def\ga{{\rm\thinspace gauss}}

\def\h0{\hbox{{\rm H}$^0$}}
\DeclareMathAlphabet{\vib}{OML}{cmm}{m}{it}
\newcommand*{\satellite}[1]{\textit{#1}}
\newcommand*{\xmm}{\satellite{XMM-Newton}}

\newcommand{\lsim}{\mathrel{\hbox{\rlap{\lower.55ex\hbox{$\sim$}} \kern-.3em \raise.4ex \hbox{$<$}}}}
\newcommand{\gsim}{\mathrel{\hbox{\rlap{\lower.55ex\hbox{$\sim$}} \kern-.3em \raise.4ex \hbox{$>$}}}}

\title[X-ray observations of the Toothbrush cluster]{Challenges to our understanding of radio relics: \\ X-ray observations of the Toothbrush cluster}
\author[G.~A.~Ogrean et al.]{G.~A.~Ogrean$^{1}$\thanks{E-mail:
gogrean@hs.uni-hamburg.de}, M. Br\"uggen$^{1}$, R.~J.~van~Weeren$^{2}$, H.~R\"ottgering$^{3}$, \and J.~H.~Croston$^{4}$, M.~Hoeft$^{5}$\\
$^{1}$Hamburger Sternwarte, Gojenbergsweg 112, 21029 Hamburg, Germany\\
$^{2}$Harvard-Smithsonian Center for Astrophysics, 60 Garden Street, Cambridge, MA 02138, USA\\
$^{3}$Leiden Observatory, Leiden University, 2300 RA Leiden, Netherlands\\
$^{4}$School of Physics and Astronomy, University of Southampton, Southampton, SO17 1BJ, UK\\
$^{5}$Th\"uringer Landessternwarte, Sternwarte 5, 07778 Tautenburg, Germany}

\begin{document}

\date{Accepted xxx xxxx xx. Received xxx xxxx xx; in original form xxx xxxx xx}

\pagerange{\pageref{firstpage}--\pageref{lastpage}} \pubyear{2013}

\maketitle

\label{firstpage}

\begin{abstract}

\noindent The cluster 1RXS J0603.3+4214 is a merging galaxy cluster that hosts three radio relics and a giant radio halo. The northern relic, the Toothbrush, is 1.9-Mpc long and has an unusual linear morphology.  According to simple diffusive shock acceleration theory, its radio spectral index indicates a Mach number of $3.3-4.6$. Here, we present results from a deep \xmm\ observation of the cluster. We observe two distinct cluster cores that have survived the merger. The presence of three shocks at or near the locations of the radio relics is confirmed by density and temperature discontinuities. However, the observation poses several puzzles that challenge our understanding of radio relics: (i) at the Toothbrush, the shock Mach number is not larger than 2, in apparent conflict with the shock strength predicted from the radio spectrum; (ii) at the Toothbrush, the shock front is, in part, spatially offset from the radio emission; (iii) at the eastern relic, we detect a temperature jump corresponding to a Mach number of approximately 2.5, but there is no associated surface brightness discontinuity. We discuss possible explanations for these findings.

\end{abstract}


\section{Introduction}
\label{s:intro}

In recent years, combined radio, X-ray, and optical observations have conclusively shown that giant radio relics -- diffuse, steep-spectrum radio objects observed in the peripheries of some clusters -- are associated with cluster mergers \citep[e.g.][]{Giacintucci2008,finoguenov10,Markevitch2010,vanWeeren2011a}. Observations, as well as cosmological simulations, suggest that relics trace shock waves triggered into the intracluster medium (ICM) during major merger events. The basic physical picture is that as shock fronts propagate through the ICM, they accelerate particles to relativistic energies and compress magnetic fields. The accelerated particles gyrate around the magnetic fields lines, consequently emitting synchrotron radiation observable at radio frequencies \citep[e.g.,][]{Feretti2012}.

The details of acceleration at merger shocks are, however, more complex and less well understood. An important question that still awaits answering is whether shocks inject thermal electrons into the cosmic-ray (CR) electron population believed to be responsible for the observed synchrotron emission, or instead they re-accelerate pre-existing CR electrons trapped in the ICM magnetic field, which itself is frozen-in to the thermal plasma. For strong shocks with Mach numbers larger than a few, as found in supernova remnants, the acceleration process, believed to be diffusive shock acceleration (DSA), is very efficient and injection dominates. But in the low Mach-number regime ($\mathcal{M}\lesssim 5$ for merger shocks), the particle acceleration efficiency is too low for simple injection to explain the observed radio brightness. At such shocks, it appears essential to consider re-acceleration of pre-existing CR electrons, as theoretical considerations and numerical simulations indicate \citep[e.g.][]{Kang2012,Pinzke2013}. Yet, observational evidence supporting this hypothesis is scarce at best.

\citet{Kang2012} simulated the particle acceleration at the relic in the galaxy cluster CIZA J2242.8+5301. They showed that models with pre-existing CRs can explain the relic by a relatively weak shock of Mach number 2. However, the injection spectral index\footnote{$\mathcal{F}\propto \nu^\alpha$} at the relic is $-0.6\pm 0.05$ \citep{vanWeeren2010}. In the DSA test particle regime, the injection spectral index relates to the shock Mach number via $\mathcal{M}^2=(2\alpha-3)/(2\alpha+1)$ \citep{Blandford1987}. Therefore, a Mach number of 2 would be much lower than the $\mathcal{M}\approx 4.5$ predicted by the radio injection spectral index. From the observational side, one way to help elucidate the nature of the accelerating particles is by comparing the radio-predicted Mach numbers of putative shocks at radio relics with the actual Mach numbers determined from X-ray observations. A current observational challenge is posed by the peripheral location of most relics, which makes the shocks difficult to detect in X-ray. Consequently, very few shocks have been detected so far at radio relics \citep{finoguenov10,Macario2011,Akamatsu2011,Ogrean2012c,Bourdin2013}. For the few shocks discovered, the X-ray-derived Mach numbers are consistent with the radio predictions. However, the small sample and the large measurement uncertainties make it impossible to draw definite conclusions about particle acceleration at merger shocks.

Here, we present results from a deep \xmm\ observation of 1RXS J0603.3+4214, a merging cluster at $z=0.225$ that hosts three radio relics and a giant radio halo \citep{vanWeeren2012}. The northern relic, the largest of the three at a length of $1.9$ Mpc, has an unusually linear morphology, with a broader part to the west; hence its nickname, the Toothbrush. The other two relics have lengths of approximately 900 and 200 kpc, and are located east and southeast of the merger \citep[see, e.g., fig. 3 of ][]{vanWeeren2012}. Numerical simulations by \citet{Brueggen2012} have shown that the linear morphology of the northern relic can be reproduced by a triple merger with mass ratios 1:1:0.07, although the simulated system failed to explain the eastern and southeastern relics. Radio observations indicate an injection spectral index at the northern relic of $-0.6$ to $-0.7$ \citep{vanWeeren2012}. Therefore, in the DSA test particle regime, the Mach number at the Toothbrush is predicted to be $\mathcal{M}=3.3-4.6$. Here, we test this prediction by locating and characterizing the shocks in the ICM of 1RXS J0603.3+4214.

The paper is organized as follows: Section \ref{s:obs} presents the observations and the data reduction. Section \ref{s:analysis} details our analysis, while in Section \ref{s:syserrors} we analyse the effect of systematic uncertainties on the spectral measurements. Sections \ref{s:shocks} and \ref{s:puzzle} discuss the results. A summary of our findings is given in Section \ref{s:conclusions}.

We assume a flat $\Lambda$CDM universe with $H_0=70$ km\,s$^{-1}$\,Mpc$^{-1}$, $\Omega_{\rm M}=0.3$, and $\Omega_{\rm \Lambda}=0.7$. At the redshift of the cluster, 1 arcmin corresponds to 217 kpc. Throughout the paper, quoted errors are $1\sigma$ statistical errors, unless stated otherwise. The normalizations of all the spectral components are given in default {\sc Xspec} units (i.e., cm$^{-5}$ for APEC components, and photons\,keV$^{-1}$\,cm$^{-2}$\,s$^{-1}$, measured at 1 keV, for power-law components) per square arcmin. All temperatures are in units of keV. Metallicities are expressed using the abundance table of \citet{angr1989}.

\section{Observations and data reduction}
\label{s:obs}

\begin{table}
\caption{Residual soft proton (ReSP) level for the MOS and pn event files, and clean exposure times. We list the ReSP calculated in the hard band, i.e. $8-12$ keV for MOS, and $10-14$ keV for pn. $\mathcal{R}_{\rm SP} < 1.2$ indicates an event file with a ReSP level that is likely negligible \citep{KuntzSnowden2008}.}
\label{tab:ReSP}
\centering
\begin{tabular}{lccc}
 \hline
     & MOS 1 & MOS 2 & pn \\ 
 \hline
   $\mathcal{R}_{\rm SP}$ & $0.98\pm 0.021$ & $0.98\pm 0.019$ & $1.14\pm 0.024$ \\
   Exp (ks) & $72.0$ & $71.7$ & $61.8$ \\ 
 \hline
\end{tabular}
\end{table}

1RXS J0603.3+4214 -- hereafter, the Toothbrush cluster -- was observed with the \xmm\ EPIC cameras on October 3-4, 2011. The observation used the medium filter, and had a total exposure time of 82 ks. We analysed the data using the \xmm\ Extended Source Analysis Software (\textsc{esas}) integrated in the Scientific Analysis System (\textsc{sas}) version 12.0.1, and the latest calibration files as of December 15, 2012. The data reduction steps are the same as those described in more detail in \citet{Ogrean2012b}. 

In summary: Raw MOS and pn event files were first filtered of soft proton flares. The filtered event files had exposure times of 72 (MOS) and 62 ks (pn). We evaluated the level of residual soft proton (ReSP) in the flare-filtered event files using the \textsc{esas}-tailored approach \citep{KuntzSnowden2008} of the algorithm presented by \citet{delucamolendi2004}. The method is based on quantifying the level of ReSP as the ratio of the count rates inside (in-FOV) and outside (out-FOV) the field of view (FOV) in the hard energy band ($8-12$ keV for MOS, $10-14$ keV for pn). The in-FOV region is chosen to be an annulus at the edge of the FOV, where there is typically little ICM emission for clusters at $z \gtrsim 0.2$. The region selection, along with the rapid drop in the detector effective area at high energies, means that the in-FOV count rate has essentially no contribution from X-ray photons. According to \citet{KuntzSnowden2008}, in-FOV/out-FOV ratios below 1.2 correspond to event files in which the ReSP level is likely negligible. The in-FOV/out-FOV ratios ($\mathcal{R}_{\rm SP}$) for each of the three flare-filtered EPIC event files are listed in Table \ref{tab:ReSP}.

The instrumental background was modelled using data from the corners (unexposed pixels) of the detectors, and filter-wheel closed datasets with hardness ratios and count rates similar to those measured during the observation. Point sources were detected using \textsc{esas}-specific routines, and excluded from the analysis. Out-of-time (OoT) events were also subtracted from the pn images and spectra. CCD \#6 of MOS1 became unoperational after a micrometeorite hit in 2005, and is automatically disregarded by \textsc{esas}.

\section{Data analysis}
\label{s:analysis}

\subsection{Imaging analysis}
\label{s:imanal}

\begin{figure}
 \begin{center}
  \includegraphics[width=\columnwidth,keepaspectratio=true,clip=true,trim=1.0cm 1.15cm 2.2cm 3cm]{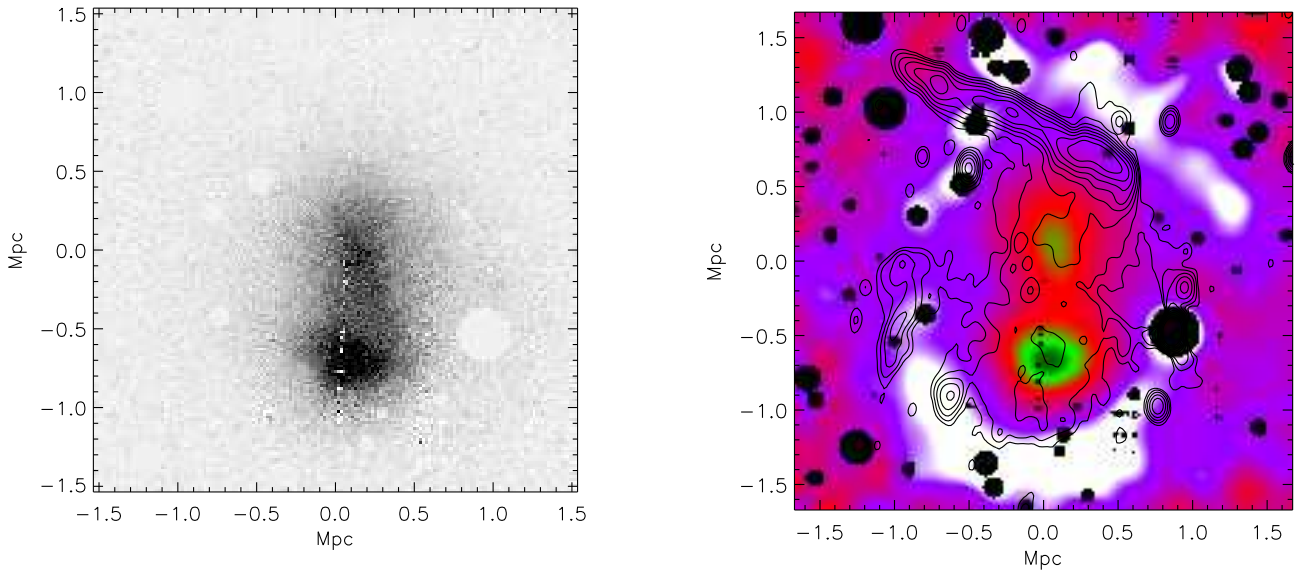}
 \end{center}
 \caption{Vignetting-corrected and instrumental background-subtracted \emph{XMM-Newton} EPIC surface brightness map of the cluster, in the energy band $0.5-4$ keV. The image was binned by a factor of 2. The gaps crossing vertically through the centre of the image are the result of overlapping MOS and pn CCD gaps.}
 \label{fig:xmmimg}
\end{figure}

\begin{figure*}
 \begin{center}
  \includegraphics[width=0.49\textwidth,keepaspectratio=true,clip=true,trim=1.0cm 1.15cm 2.2cm 3cm]{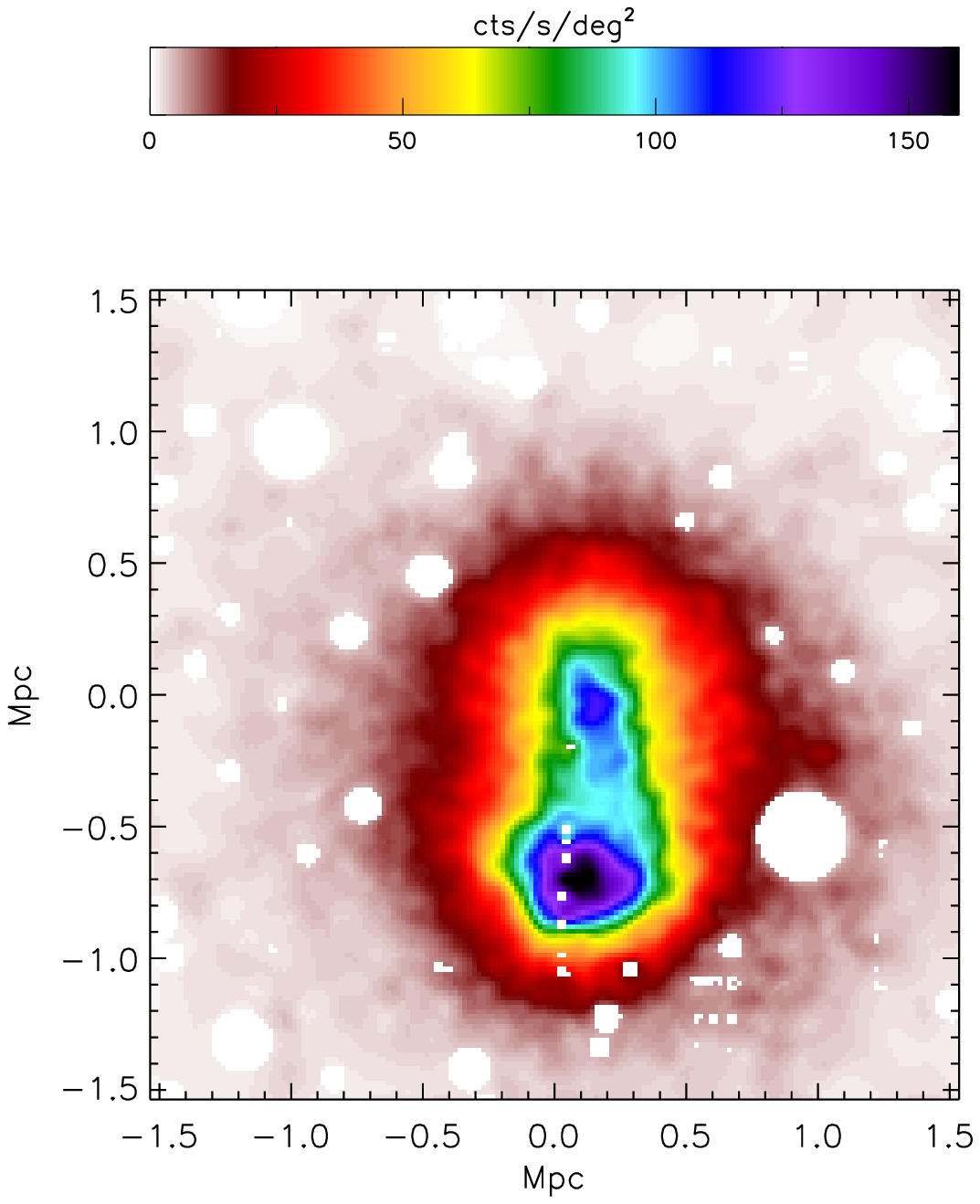}
  \includegraphics[width=0.49\textwidth,keepaspectratio=true,clip=true,trim=1.0cm 1.15cm 2.2cm 3cm]{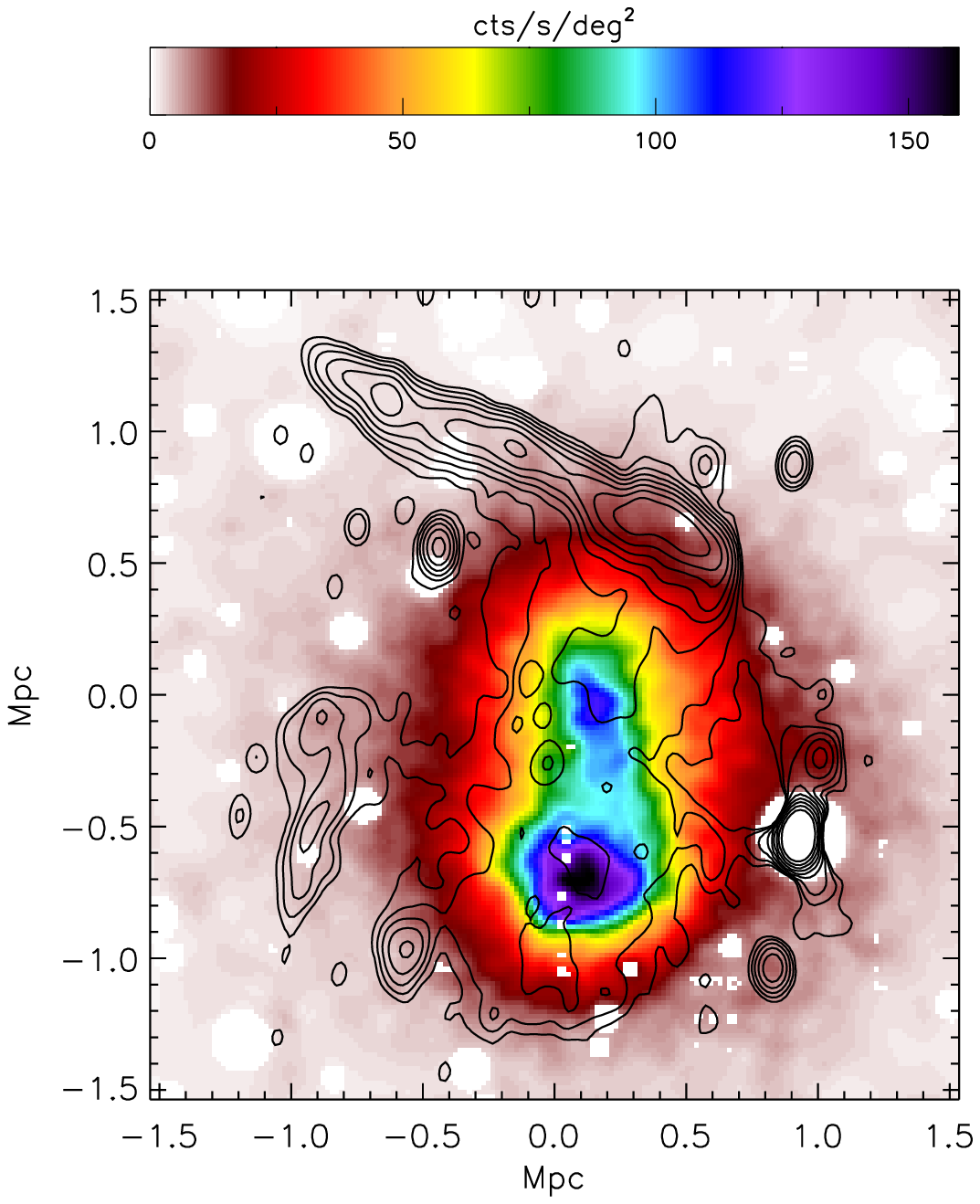}
 \end{center}
 \caption{\emph{Left:} Adaptively-smoothed, vignetting-corrected, and instrumental background-subtracted EPIC surface brightness map, in the energy band $0.5-4$ keV. The adaptive smoothing had a target SNR of 5. \emph{Right:} Same image, with overlaid WSRT radio contours. Contours are drawn at $[1,2,4,...]\times 120\; \mu$Jy/beam, using a combined radio brightness map created from images at wavelengths 18, 21, and 25 cm.}
 \label{fig:xmmimg-smooth}
\end{figure*}

\begin{figure}
 \begin{center}
  \includegraphics[width=0.49\textwidth,keepaspectratio=true,clip=true,trim=1.0cm 1.15cm 2.2cm 3cm]{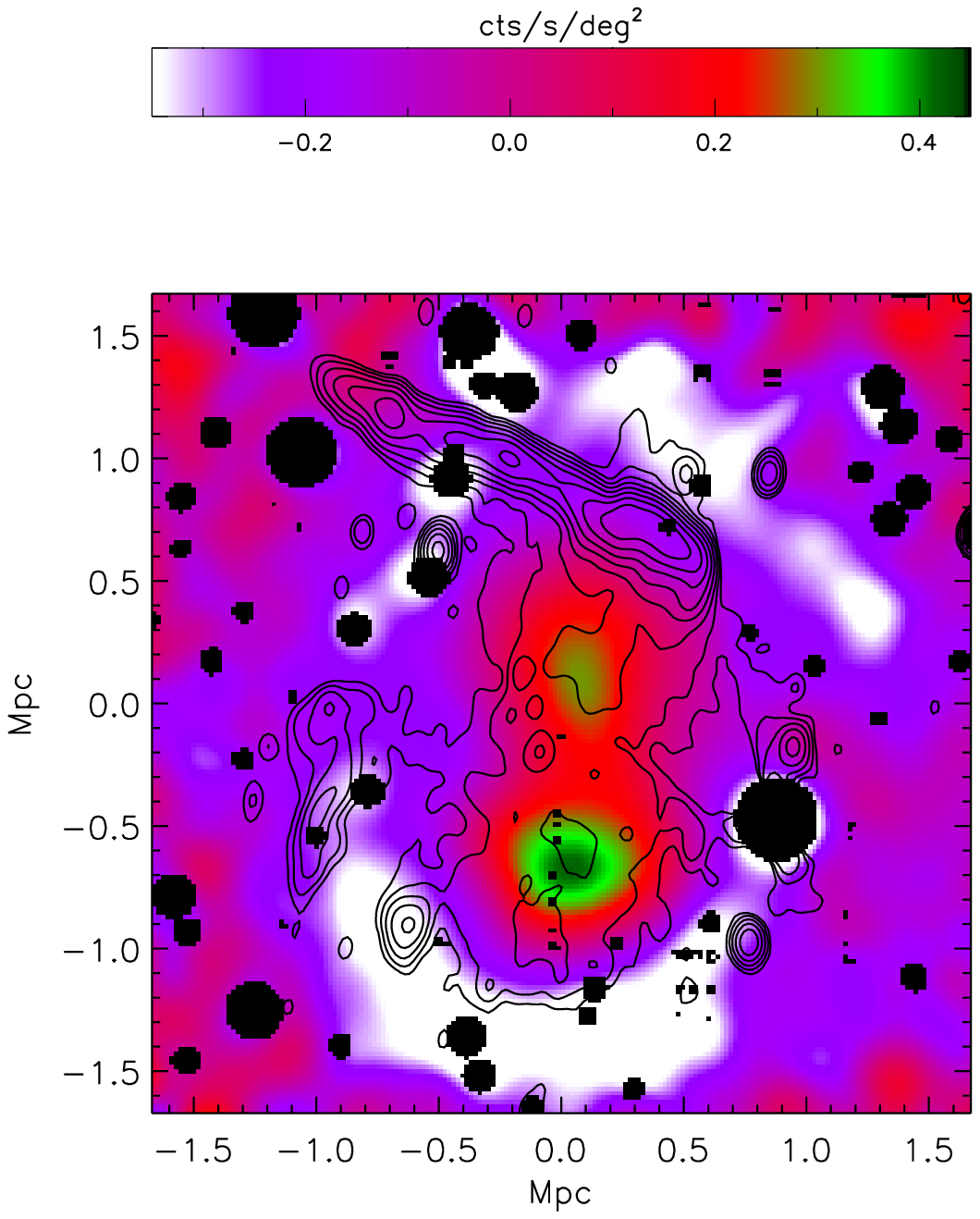}
 \end{center}
 \caption{Unsharp-masked image emphasizing the surface brightness discontinuities. Overlaid are the same \emph{WSRT} radio contours as in Figure \ref{fig:xmmimg-smooth}. Point sources and zero-exposure pixels are shown in black, to distinguish between shock features (in white) and artifacts around no-flux regions.}
 \label{fig:unsharp-mask-shocks}
\end{figure}

Figure \ref{fig:xmmimg} presents the EPIC $0.5-4$ keV surface brightness map of the Toothbrush cluster. The map was instrumental background-subtracted, vignetting-corrected, and binned by 2. The sub-band $1.2-1.9$ keV, which is dominated by fluorescent instrumental lines (Al K$\alpha \sim 1.49$ keV, Si K$\alpha \sim 1.75$ keV), was not included in the image. The null pixels passing in a straight vertical line through the centre of the image are caused by the superposition of CCD gaps in the three detectors. Figure \ref{fig:xmmimg-smooth} shows the EPIC adaptively smoothed surface brightness map in the same energy band. Two distinct cluster cores are evident in the maps. Both of them have been significantly disturbed by the merger. A bridge connects the northern, less massive core, to the northwestern part of the southern core. The bridge is brighter in the north, closer to the smaller core.

Surface brightness discontinuities are present south of the more massive core, and NW of the subcore. They are more clearly seen in the unsharp-masked image in Figure \ref{fig:unsharp-mask-shocks}. The unsharp-masked images were created using two $0.5-4$ keV surface brightness maps smoothed with Gaussians of widths 25 and 100 arcsec, by dividing their difference to their sum. The large smoothing scale of 100 arcsec removes substructure from the image, while preserving the global X-ray morphology. The smaller smoothing scale of 25 arcsec surpresses Poissonian noise, and reduces small-scale substructure and pixel-to-pixel variations (e.g., introduced by the OoT events correction, CCD gaps). We are searching for surface brightness edges on scales of $5-10$ arcmin (the Toothbrush relic has a length of almost 9 arcmin), much larger than the smoothing scales used. We tested the robustness of our unsharp-masked images by experimenting with different smoothing scales and unsharp-masking procedures \citep[e.g.,][]{Sanders2012}. The enhanced surface brightness edges are visible for various smoothing scales in the range of $10-200$ arcsec, but our choices of $25$ and $100$ arcsec are the best compromise between making the edges clear and removing substructure. We point out that the same unsharp-masking technique has been used by \citet{Russell2010} to highlight substructure and surface brightness discontinuities in Abell 2146. The white regions in the map indicate shock features, except when they surround excluded point sources or zero-exposure pixels (shown in black). The fact that the shocks are asymmetric with respect to the merger axis (the imaginary line connecting the two cluster cores) indicates a merger with non-zero impact parameter. Interestingly, the NW shock, which was expected to trace the Toothbrush, is visible only along the broadest part of the relic, and extends about 700 kpc further to the SW, beyond the radio emission.

\emph{In summary, the surface brightness map shows evidence of a merging galaxy cluster involving at least two clusters. Two cluster cores have survived the merger, and are connected by a dense plasma bridge. Shocks triggered during the merger are observed as surface brightness discontinuities N-NW and S-SE of the northern and southern cores, respectively. The orientation of the shocks with respect to the merger axis suggests a non-zero impact parameter. The N-NW shock is visible only near the broadest part of the Toothbrush, and extends more than 0.5 Mpc beyond it, towards the SW.}

\subsection{Temperature distribution}
\label{s:temperaturedistribution}

\begin{figure*}
 \begin{center}
  \includegraphics[width=0.49\textwidth,keepaspectratio=true,clip=true,trim=0.5cm 0.2cm 1.1cm 1.8cm]{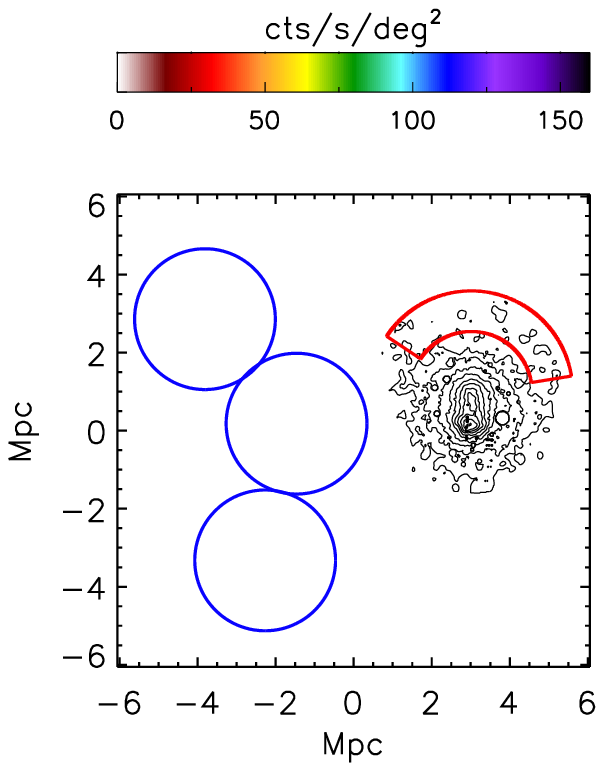}
  \includegraphics[width=0.49\textwidth,keepaspectratio=true,clip=true,trim=0.5cm 0.2cm 1.1cm 1.8cm]{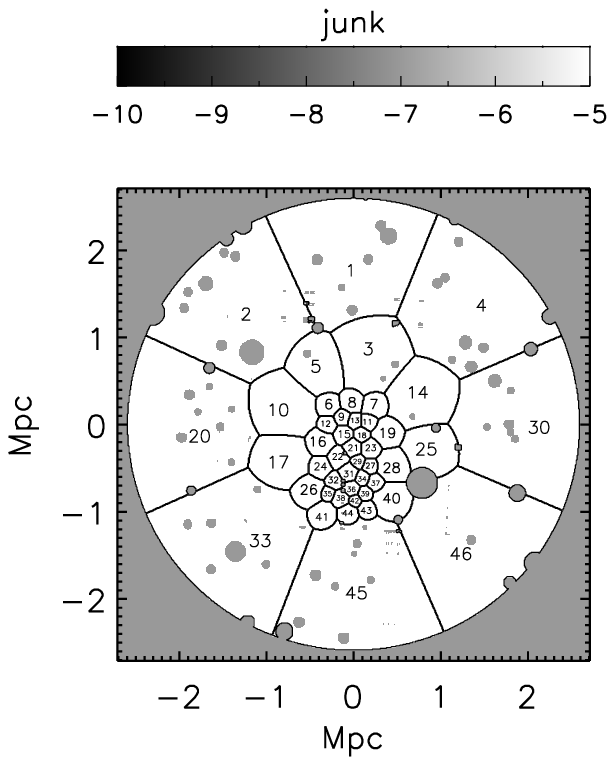}
 \end{center}
 \caption{\emph{Left:} Background regions. The \xmm\ region is shown in red, while the ROSAT regions are shown in blue. Contours are X-ray surface brightness contours. \emph{Right:} Bin number map. Each bin has roughly 3600 ICM plus sky background counts. Gray areas correspond to detected point-like sources and to regions outside the central 24 arcmin.}
 \label{fig:regions}
\end{figure*}

\begin{figure*}
 \begin{center}
  \includegraphics[width=0.495\textwidth,keepaspectratio=true,trim=1.0cm 1.15cm 2.2cm 0cm]{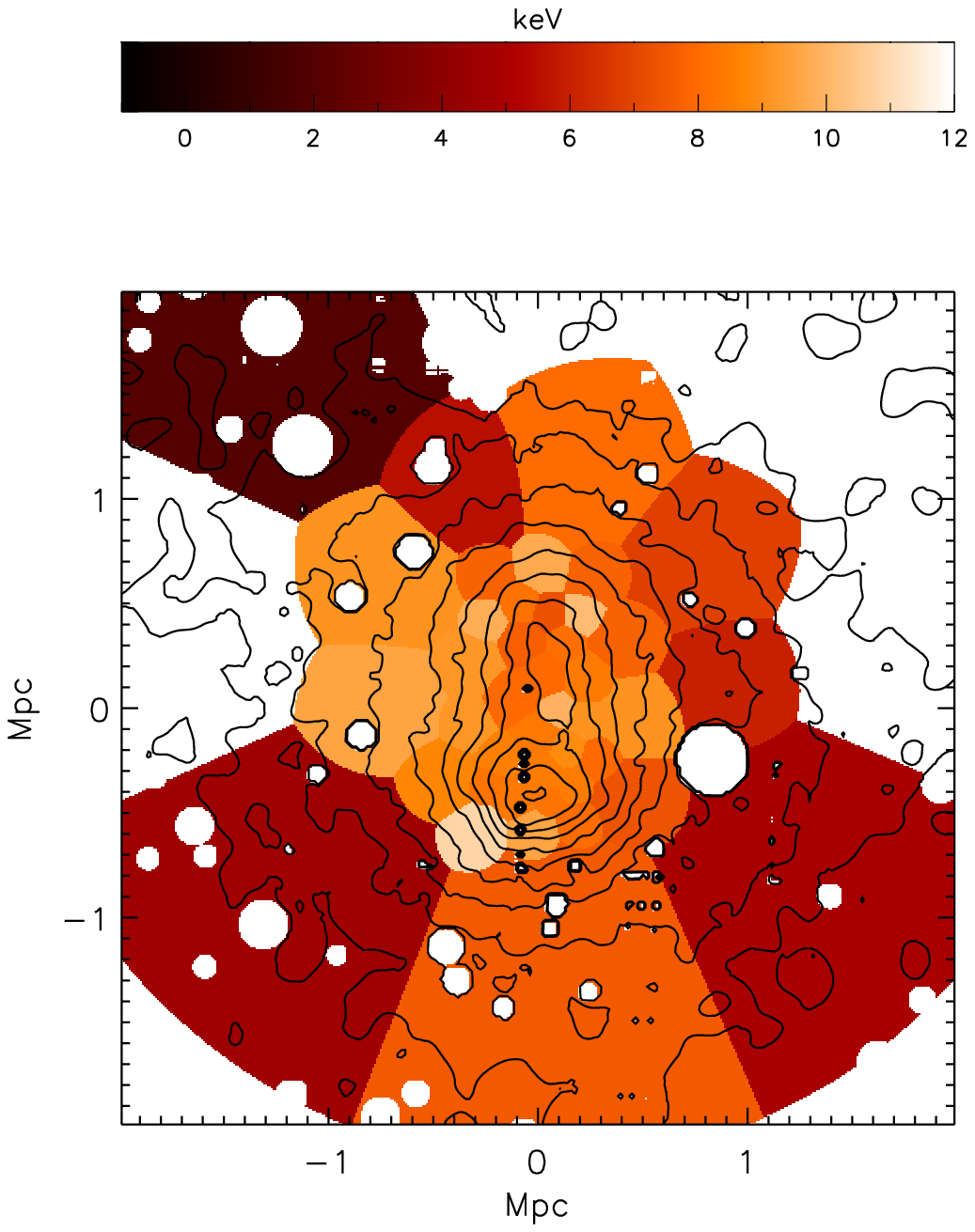}
  \includegraphics[width=0.495\textwidth,keepaspectratio=true,trim=1.0cm 1.15cm 2.2cm 0cm]{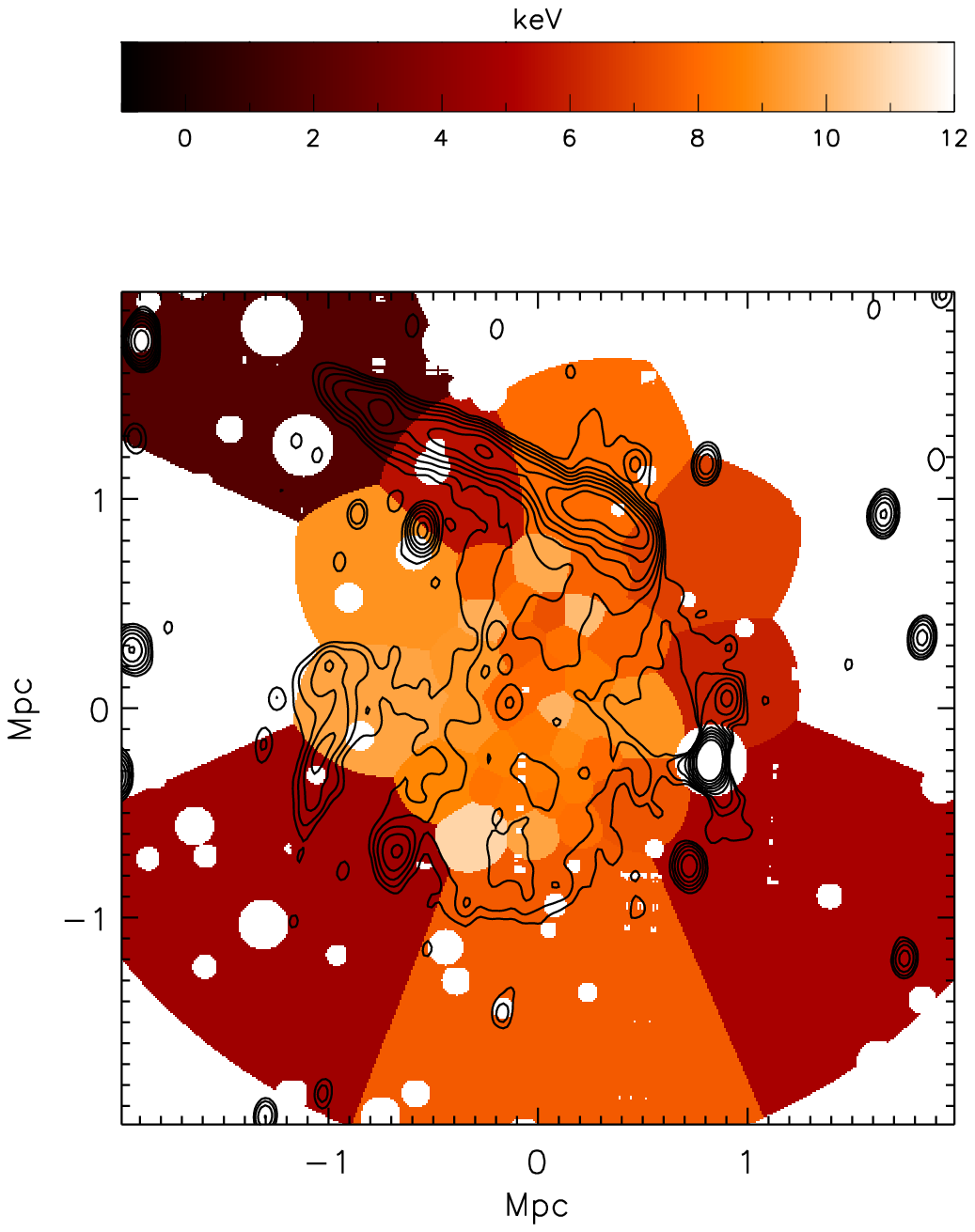}
 \end{center}
 \caption{Temperature map of the cluster. Overlaid are X-ray contours (left) and radio contours (right). Each bin has approximately 3600 counts from source plus sky-background.}
 \label{fig:temperature-map-xray}
\end{figure*}

\begin{table}
\caption{Best-fit sky background model. Frozen parameters are marked with a dagger ($\dagger$). All parameters are expressed in default {\sc XSpec} units. The normalizations were scaled to correspond to an area of 1 sq arcmin.}
\label{tab:bkg}
\centering
\begin{tabular}{lccc}
 \hline
   Component & $T_{\rm X}$ & $\Gamma$ & $\mathcal{N}_{\rm X}$ \\ 
 \hline
   LHB & $0.08^{\dagger}$ & -- & $(4.1\pm 0.75) \times 10^{-7}$ \\
   GH  & $0.18_{-0.0030}^{+0.0032}$ & -- & $(2.2\pm 0.086) \times 10^{-6}$ \\
   HF  & $0.97_{-0.052}^{+0.050}$ & -- & $(1.7\pm 0.21) \times 10^{-7}$ \\
   CXB & -- & $1.41^{\dagger}$ & $(6.5\pm 0.30) \times 10^{-7}$ \\
 \hline
\end{tabular}
\end{table}

\begin{table}
\caption{Best-fit ICM temperatures and normalizations for the binned regions in Figure \ref{fig:regions}. We only list bins for which both the temperature and the normalization were constrained. The metallicity was frozen to 0.2 solar, using the abundances table of \citet{angr1989} and the photoelectric absorption cross-sections of \citet{Verner1996}. Values are given in default {\sc Xspec} units. Normalizations are scaled to an area of 1 sq arcmin.}
\label{tab:icm}
\centering
\begin{tabular}{lcc}
 \hline
   Bin & $T_{\rm X}$ & $\mathcal{N}_{\rm X}$ \\ 
 \hline
   2 & $2.0_{-0.44}^{+0.64}$ & $2.1_{-0.40}^{+0.41} \times 10^{-6}$ \\
   3 & $8.0_{-0.91}^{+1.1}$  & $2.2_{-0.065}^{+0.066} \times 10^{-5}$ \\
   5 & $5.6_{-0.50}^{+0.63}$ & $2.9_{-0.10}^{+0.12} \times 10^{-5}$ \\
   6 & $7.8_{-0.73}^{+0.75}$ & $1.3_{-0.035}^{+0.036} \times 10^{-4}$ \\
   7 & $7.8_{-0.67}^{+0.69}$ & $1.5_{-0.038}^{+0.039} \times 10^{-4}$ \\
   8 & $9.7_{-0.94}^{+0.95}$ & $1.8_{-0.035}^{+0.036} \times 10^{-4}$ \\
   9 & $8.2_{-0.75}^{+1.0}$  & $3.3_{-0.091}^{+0.093} \times 10^{-4}$ \\
  10 & $9.1\pm 1.2$          & $3.2_{-0.075}^{+0.093} \times 10^{-5}$ \\
  11 & $10.2_{-1.2}^{+1.9}$  & $2.9_{-0.073}^{+0.075} \times 10^{-4}$ \\
  12 & $9.8_{-1.2}^{+1.3}$   & $2.1_{-0.055}^{+0.056} \times 10^{-4}$ \\
  13 & $7.3_{-0.60}^{+0.67}$ & $(4.2\pm 0.11) \times 10^{-4}$ \\
  14 & $6.9_{-0.51}^{+0.85}$ & $2.6_{-0.075}^{+0.076} \times 10^{-5}$ \\
  15 & $7.7_{-0.54}^{+0.56}$ & $4.4_{-0.093}^{+0.095} \times 10^{-4}$ \\
  16 & $9.2_{-0.82}^{+0.91}$ & $(1.9\pm 0.037) \times 10^{-4}$ \\
  17 & $9.6\pm 1.1$          & $(4.6\pm 0.11) \times 10^{-5}$ \\
  18 & $8.1_{-0.65}^{+0.73}$ & $(4.6\pm 0.11) \times 10^{-4}$ \\
  19 & $7.8_{-0.54}^{+0.55}$ & $1.9_{-0.037}^{+0.038} \times 10^{-4}$ \\
  21 & $8.0_{-0.58}^{+0.60}$ & $(5.5\pm 0.12) \times 10^{-4}$ \\
  22 & $8.0_{-0.55}^{+0.57}$ & $3.7_{-0.079}^{+0.080} \times 10^{-4}$ \\
  23 & $8.4_{-0.66}^{+1.3}$  & $3.4_{-0.093}^{+0.086} \times 10^{-4}$ \\
  24 & $9.2_{-0.91}^{+1.1}$  & $1.9_{-0.045}^{+0.046} \times 10^{-4}$ \\
  25 & $6.0_{-0.52}^{+0.54}$ & $(4.8\pm 0.15) \times 10^{-5}$ \\
  26 & $8.8_{-0.76}^{+1.0}$  & $1.5_{-0.034}^{+0.040} \times 10^{-4}$ \\
  27 & $9.1_{-0.90}^{+1.2}$  & $3.8_{-0.091}^{+0.094} \times 10^{-4}$ \\
  28 & $9.1_{-0.95}^{+0.96}$ & $1.5_{-0.032}^{+0.038} \times 10^{-4}$ \\
  29 & $10.0_{-1.1}^{+1.3}$  & $(5.1\pm 0.12) \times 10^{-4}$ \\
  31 & $8.4_{-0.63}^{+0.92}$ & $(5.0\pm 0.12) \times 10^{-4}$ \\
  32 & $8.6_{-0.64}^{+0.93}$ & $4.7_{-0.099}^{+0.12} \times 10^{-4}$ \\
  33 & $4.6_{-0.73}^{+0.94}$ & $6.4_{-0.44}^{+0.46} \times 10^{-6}$ \\
  34 & $8.9_{-0.76}^{+1.1}$  & $(4.6\pm 0.11) \times 10^{-4}$ \\
  35 & $8.3_{-0.78}^{+1.1}$  & $4.4_{-0.13}^{+0.14} \times 10^{-4}$ \\
  36 & $8.5_{-0.60}^{+0.78}$ & $6.5_{-0.13}^{+0.15} \times 10^{-4}$ \\
  37 & $8.0_{-0.74}^{+0.82}$ & $3.5_{-0.095}^{+0.097} \times 10^{-4}$ \\
  38 & $8.4_{-0.57}^{+0.68}$ & $6.8_{-0.13}^{+0.15} \times 10^{-4}$ \\
  39 & $8.5_{-0.70}^{+1.3}$  & $5.2_{-0.15}^{+0.14} \times 10^{-4}$ \\
  40 & $7.4_{-0.56}^{+0.58}$ & $(1.5\pm 0.033) \times 10^{-4}$ \\
  41 & $10.9_{-1.0}^{+1.6}$  & $1.7_{-0.031}^{+0.038} \times 10^{-4}$ \\
  42 & $8.3_{-0.71}^{+0.93}$ & $6.4_{-0.17}^{+0.18} \times 10^{-4}$ \\
  43 & $8.1_{-0.68}^{+0.88}$ & $3.4_{-0.087}^{+0.089} \times 10^{-4}$ \\
  44 & $9.6_{-0.93}^{+0.94}$ & $4.1_{-0.081}^{+0.082} \times 10^{-4}$ \\
  45 & $7.6_{-0.88}^{+0.91}$ & $1.7_{-0.047}^{+0.048} \times 10^{-5}$ \\
  46 & $5.0_{-0.77}^{+1.2}$  & $(8.6\pm 0.52) \times 10^{-6}$ \\
 \hline
\end{tabular}
\end{table}

We used the weighted Voronoi tesselations (WVT) binning algorithm of \citet{DiehlStatler2006} to bin the cluster image in regions of 3600 counts from source plus sky background, starting from the centre of the FOV. We then extracted spectra and response files corresponding to each region, and binned the spectra to a minimum of 30 counts per bin. Similarly, we also extracted global spectra from a circular region encompassing most of the cluster emission. Modelling the ICM spectra requires a knowledge of the sky background. Therefore, sky background spectra were extracted from a partial annulus at the edge of the FOV, where the surface brightness is the lowest; this region is least likely to contain ICM emission. Like the source spectra, the sky background spectra were also binned to a minimum of 30 counts per bin. To contrain better the low-energy background components, we also used three \emph{ROSAT} All-Sky Survey (RASS) spectra extracted from regions near the cluster. The sky background regions and the binned cluster map are shown in Figure \ref{fig:regions}.

We fitted all the spectra in parallel using {\sc XSpec} v. 12.7.1. Before fitting, the corresponding instrumental background spectra were subtracted from the source and sky background spectra. The sky background was modelled as the sum of Local Hot Bubble (LHB), Galactic Halo (GH), Hot Foreground\footnote{The Toothbrush cluster is located at a Galactic latitude of $\approx 9.7$ degrees, where hot foreground emission is possibly present.} \citep[HF; e.g.,][]{Simionescu2011}, and Cosmic X-ray Background (CXB) emission. For the thermal components (LHB, GH, HF), the abundances were fixed to solar and the redshifts to zero, as customary. The LHB temperature was fixed to 0.08 keV \citep{Sidher1996,KuntzSnowden2000}. The CXB was modelled as a power-law component, with a frozen index of 1.41 \citep{delucamolendi2004}. Because no point sources were subtracted from the RASS spectra, the RASS CXB normalization was fixed to $8.85\times 10^{-7}$ ${\rm photons\,\,keV^{-1}\,\,cm^{-2}\,\,s^{-1}\,\,arcmin^{-2}}$ \citep{Moretti2003}. For describing ICM emission, we added to the sky background model a single-temperature thermal component with a fixed redshift of 0.225 and a fixed metallicity of 0.2 solar \citep[e.g.,][]{LeccardiMolendi2008b}. The absorbed components, i.e. all except the LHB, were multiplied by a photoelectric absorption model with a free X-ray column density; an exception are the background spectra, for which the column density was fixed to $2.14\times 10^{21}$ cm$^{-2}$ -- the Galactic H{\sc i} column density listed in the Leiden/Argentine/Bonn (LAB) Survey \citep{Kalberla2005} in a 0.5-degree circle around the cluster -- because of degeneracy between the background model normalizations and the absorption. The fluorescent instrumental lines of Al K$\alpha$ ($E\approx 1.49$ keV) and Si K$\alpha$ ($E\approx 1.74$ keV) vary in intensity with time and across each of the EPIC detectors, and therefore they are not included in the instrumental background model. We modelled them separately using two zero-width Gaussian components with free normalizations, but with central energies coupled between spectra.

The best-fit background model is summarized in Table \ref{tab:bkg}. Table \ref{tab:icm} shows the ICM temperatures and normalizations for all binned regions for which we were able to constrain the spectral parameters. The fit had a $\chi^2/{\rm d.o.f.}=12380.43/12110$. Figure \ref{fig:temperature-map-xray} presents the resulting temperature map, with overlaid X-ray and radio contours.

\emph{The cluster has an average temperature of $7.8\pm 0.13$ keV, and $L_{\rm X,\, 0.5-7\, keV}=1.9\times 10^{45}$ erg~s$^{-1}$ ($L_{\rm X,\, 0.1-2.4 keV}=7.7\times 10^{44}$ erg~s$^{-1}$). Along the Toothbrush, the temperature decreases from approximately 8~keV to the SW, to about 2~keV to the NE. E-SE of the merger, there is a region of hot plasma with temperatures of around 11~keV. For the bins shown in Figure \ref{fig:temperature-map-xray}, the average temperature is 8.1~keV, with a standard deviation of 1.6~keV. Most temperatures are within $1\sigma$ of the average when statistical errors are considered.}

\section{Systematic uncertainties}
\label{s:syserrors}

\begin{figure*}
 \begin{center}
  \includegraphics[width=1\textwidth,keepaspectratio=true]{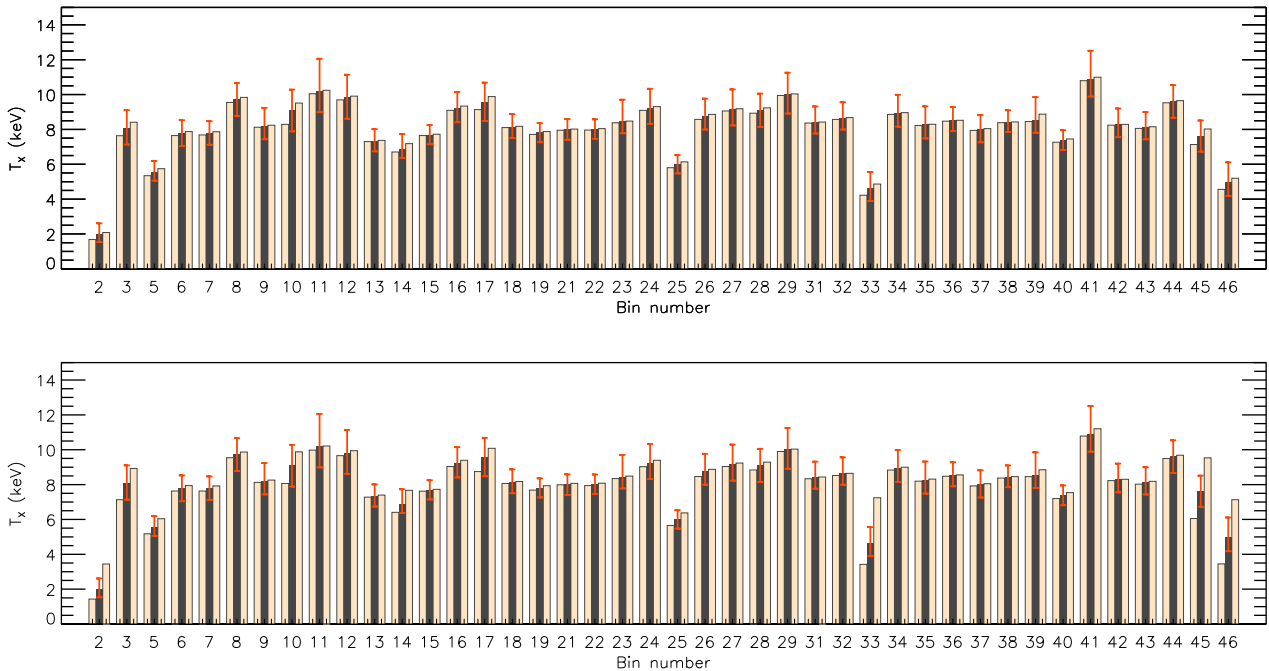}
 \end{center}
 \caption{\emph{Top:} Statistical and systematic errors on the best-fit temperature measurements. Charcoal bars show the best-temperatures listed in Table \ref{tab:icm}, while bisque bars show the minimum and maximum temperatures when systematic uncertainties on the sky background parameters are taken into account. \emph{Bottom:} ICM temperature uncertainties introduced by a $5\%$ variation in the normalization of the instrumental background. Charcoal bars show the best-temperatures listed in Table \ref{tab:icm}, while bisque bars show the minimum and maximum temperatures obtained when the instrumental background level is varied by $\pm 5\%$. In both plots, error bars are $1\sigma$ statistical errors.}
 \label{fig:syserr}
\end{figure*}

\begin{figure}
 \begin{center}
  \includegraphics[width=1\columnwidth,keepaspectratio=true,trim=1.0cm 1.15cm 2.2cm 0cm]{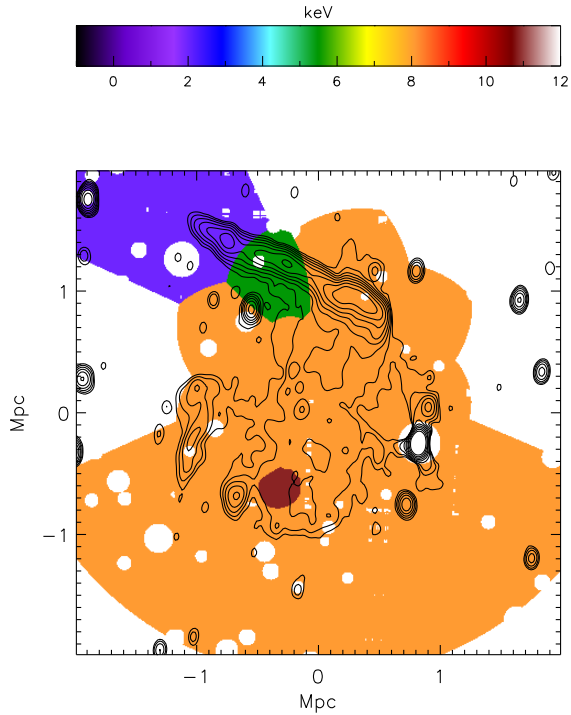}
 \end{center}
 \caption{Temperature map, with statistical and systematic errors taken into account. All bins with temperatures within $1\sigma$ of the average are shown in orange. Overlaid are the same radio contours as in Figure \ref{fig:xmmimg-smooth}.}
 \label{fig:huub-map}
\end{figure}

There are three sources of systematic errors on the measured ICM parameters: the X-ray column density in the sky background region, systematic uncertainties on the sky background parameters, and uncertainties on the instrumental background model.

In a circular region of radius 1-degree around the cluster, there are seven Galactic H{\sc i} measurements, with corresponding Galactic hydrogen column densities in the range $(1.86-2.47)\times 10^{21}$~cm$^{-2}$. The weighted average column density is $2.15\times 10^{21}$~cm$^{-2}$, essentially equal to the column density measurement at the position closest to the cluster. To analyze the effect of an uncertain X-ray column density in the sky background region, we refitted all the spectra, once with the sky background column density fixed to $1.86\times 10^{21}$~cm$^{-2}$ (the lowest $N_{\rm H}$ in the region), and once with it fixed to $2.47\times 10^{21}$~cm$^{-2}$ (the highest $N_{\rm H}$ value). Using each of the two new sets of best-fit background parameters and the corresponding statistical errors, we refitted the ICM spectra while varying (one by one) the sky background parameters by their $1\sigma$ statistical errors; the sky background model was fixed in these fits. The resulting minimum and maximum ICM parameters for each bin define the ranges for these parameters when systematic uncertainties on the sky background parameters and column density are taken into account. 

Figure \ref{fig:syserr} shows the best-fit temperatures listed in Table \ref{tab:icm}, with overlaid $1\sigma$ statistical error bars, and the ranges of systematic uncertainties. The statistical errors dominate, so $N_{\rm H}$ uncertainties and $1\sigma$ variations on any of the background parameters have only a small effect on the best-fit ICM temperatures. The same is also true for normalizations (not shown in Figure \ref{fig:syserr}), for which the effect is even stronger. 

We evaluated separately the effect of a possible $5\%$ error on the normalization of the quiescent particle background. For each bin spectrum (see Figure \ref{fig:regions}), the normalizations of the MOS and pn spectra were simultaneously increased and then decreased by $5\%$. The sky background was fixed to the model summarized in Table \ref{tab:bkg}, and the spectra were fitted again to obtain new best-fit ICM temperature and normalization, and X-ray column density. The minimum and maximum ICM parameters yielded by these fits are shown graphically in Figure \ref{fig:syserr}. For most bins, the temperature uncertainty introduced by a potential $5\%$ error on the QPB normalization is smaller than the statistical error. Exceptions are the outer bins (bins 2, 33, 45, and 46), for which $5\%$ changes in the instrumental background level resulted in possible temperature ranges significantly larger than $2\sigma$.

In Figure \ref{fig:huub-map} we show again the temperature map, including the uncertainties introduced by statistical and systematic errors. Bins with temperature ranges consistent, within $1\sigma$, with the average temperature of 8.1~keV are all shown in orange. Bins 2, 5, and 41, which have temperatures markedly different from the average, are shown in purple, green, and dark red, respectively.

\section{Shocks}
\label{s:shocks}

\begin{figure*}
 \begin{center}
  \includegraphics[width=0.49\textwidth,keepaspectratio=true,clip=true,trim=1.5cm 1.15cm 2.2cm 3cm]{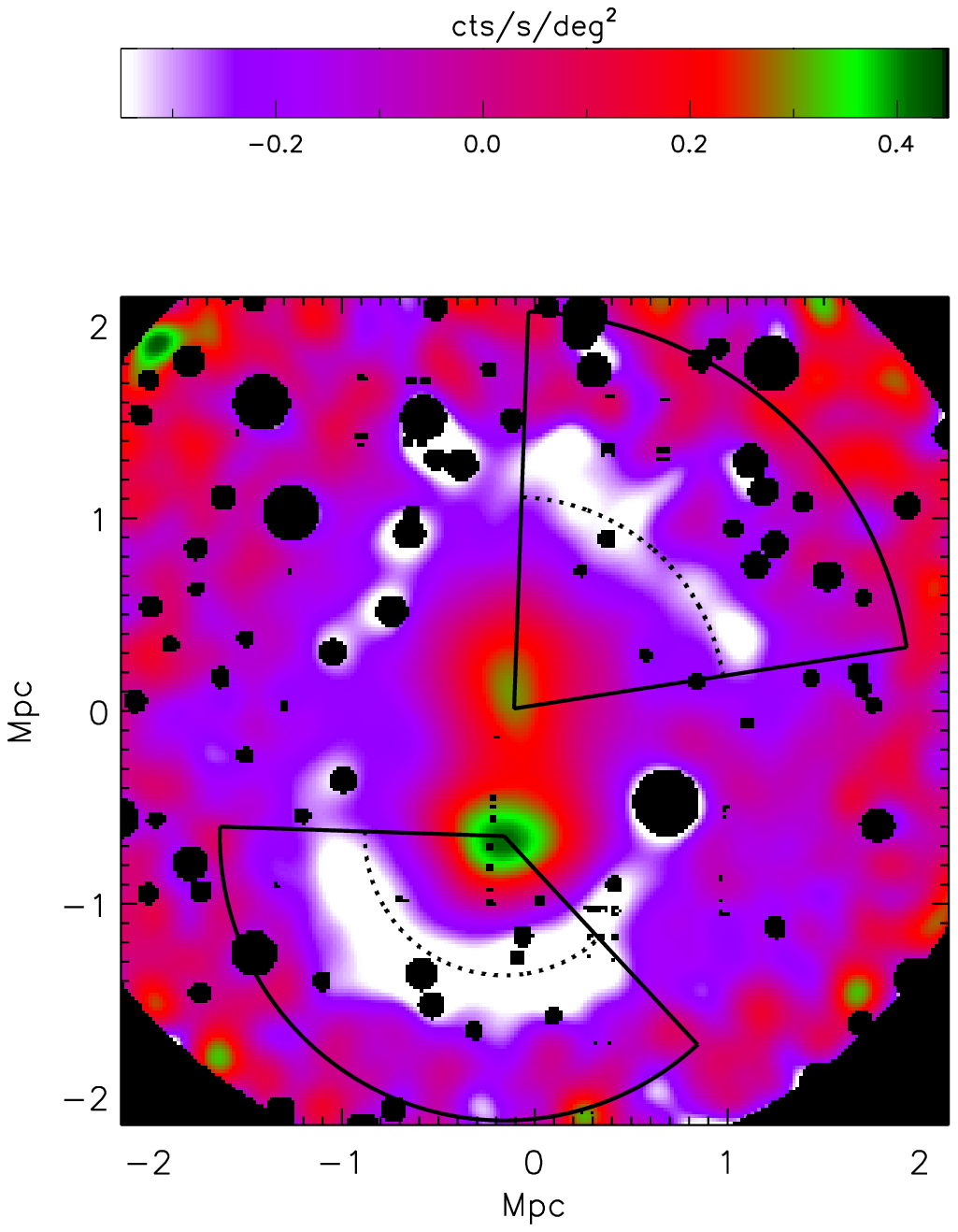}
  \includegraphics[width=0.49\textwidth,keepaspectratio=true,clip=true,trim=1.5cm 1.15cm 2.2cm 3cm]{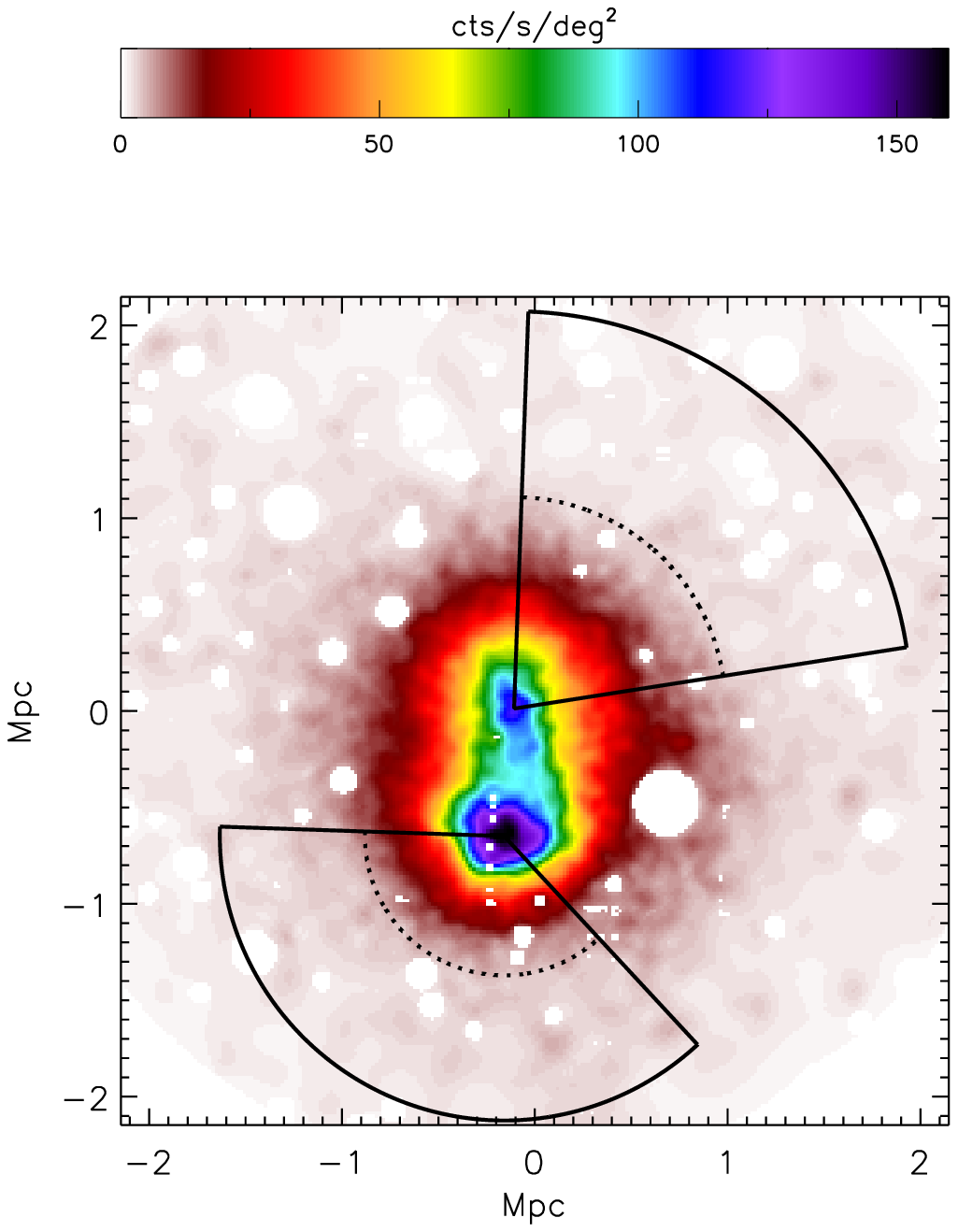}
 \end{center}
 \caption{The same unsharp-masked and adaptively smoothed images as shown in Figures \ref{fig:xmmimg-smooth} and \ref{fig:unsharp-mask-shocks}, with overlaid regions showing the sectors used for fitting the surface brightness discontinuities. Dotted lines show the best-fit positions of the density discontinuities. The surface brightness profiles and best-fit models are shown in Figure \ref{fig:shock-profiles}.}
 \label{fig:shock-pies}
\end{figure*}

\begin{figure*}
 \begin{center}
  \includegraphics[width=0.49\textwidth,keepaspectratio=true,clip=true]{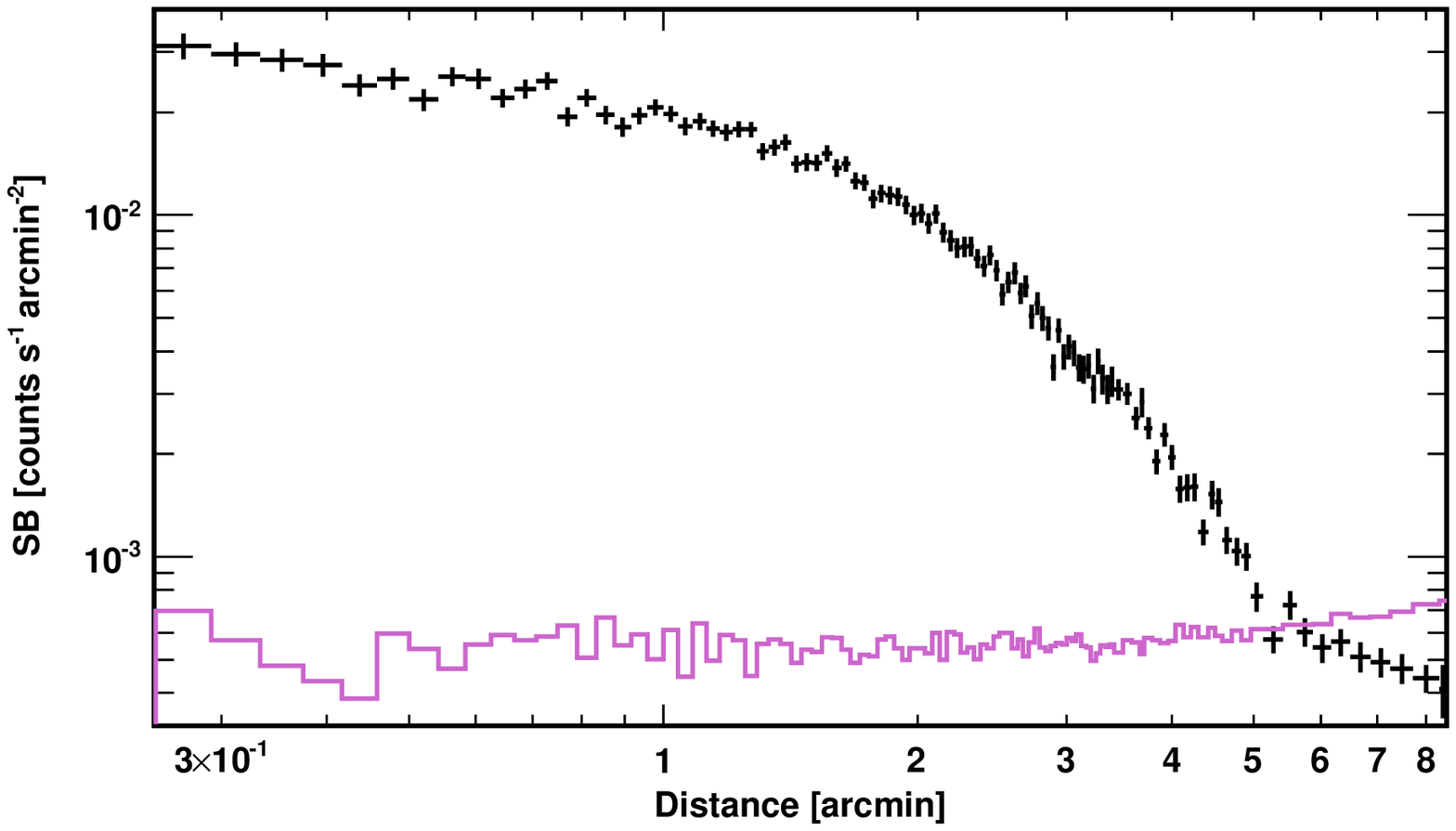}
  \includegraphics[width=0.49\textwidth,keepaspectratio=true,clip=true]{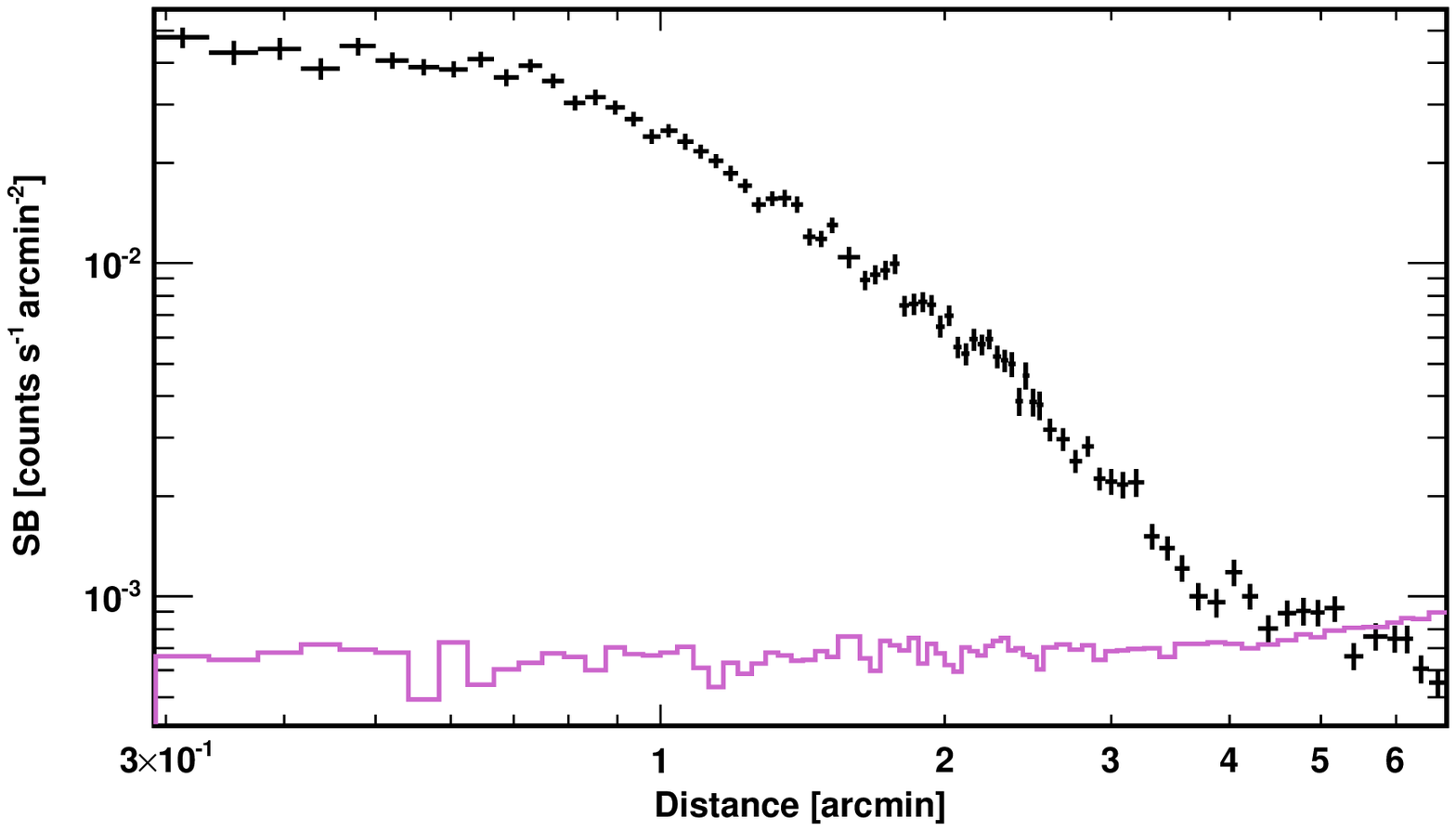}
  \includegraphics[width=0.497\textwidth,keepaspectratio=true,clip=true]{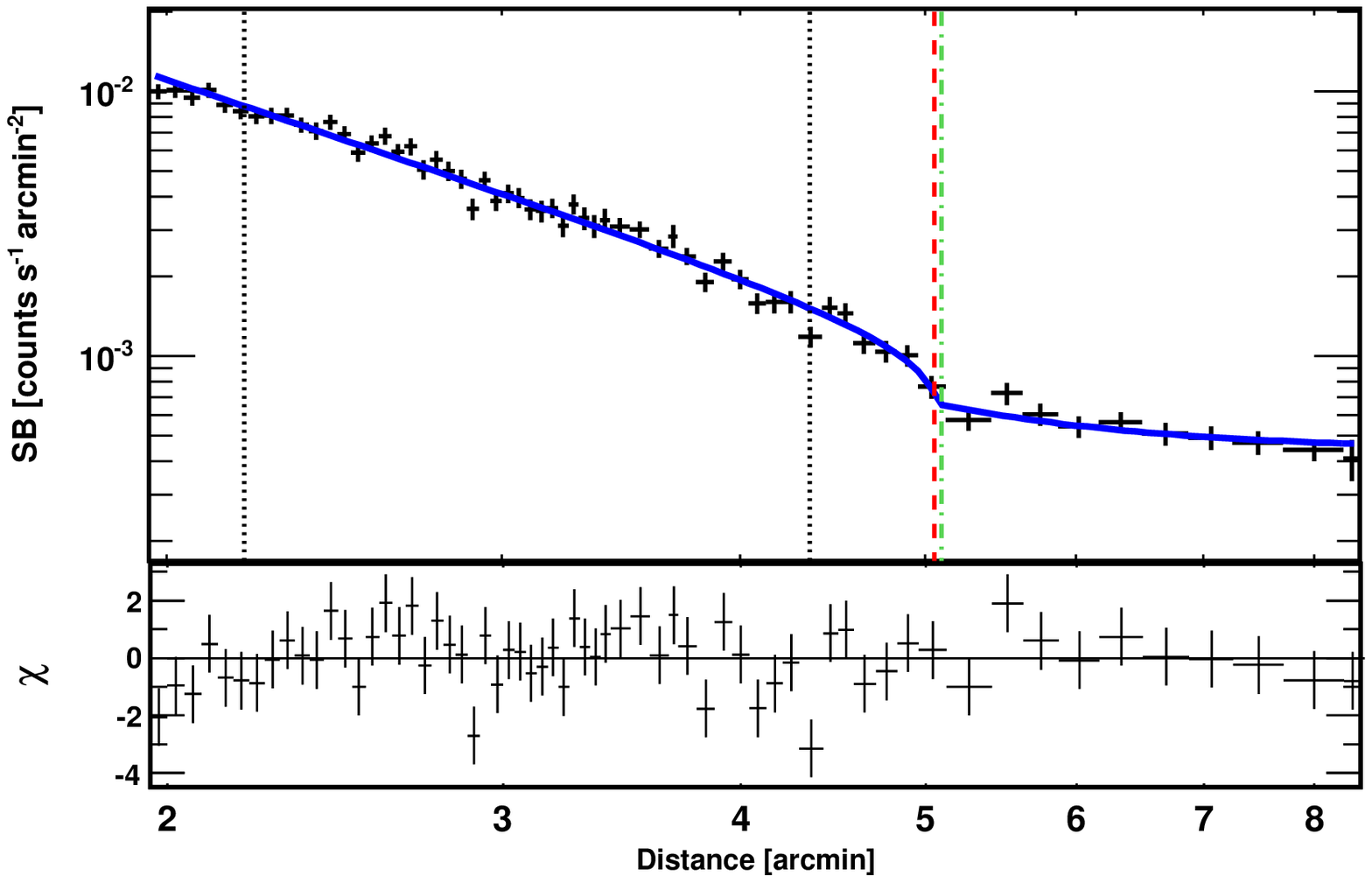}
  \includegraphics[width=0.497\textwidth,keepaspectratio=true,clip=true]{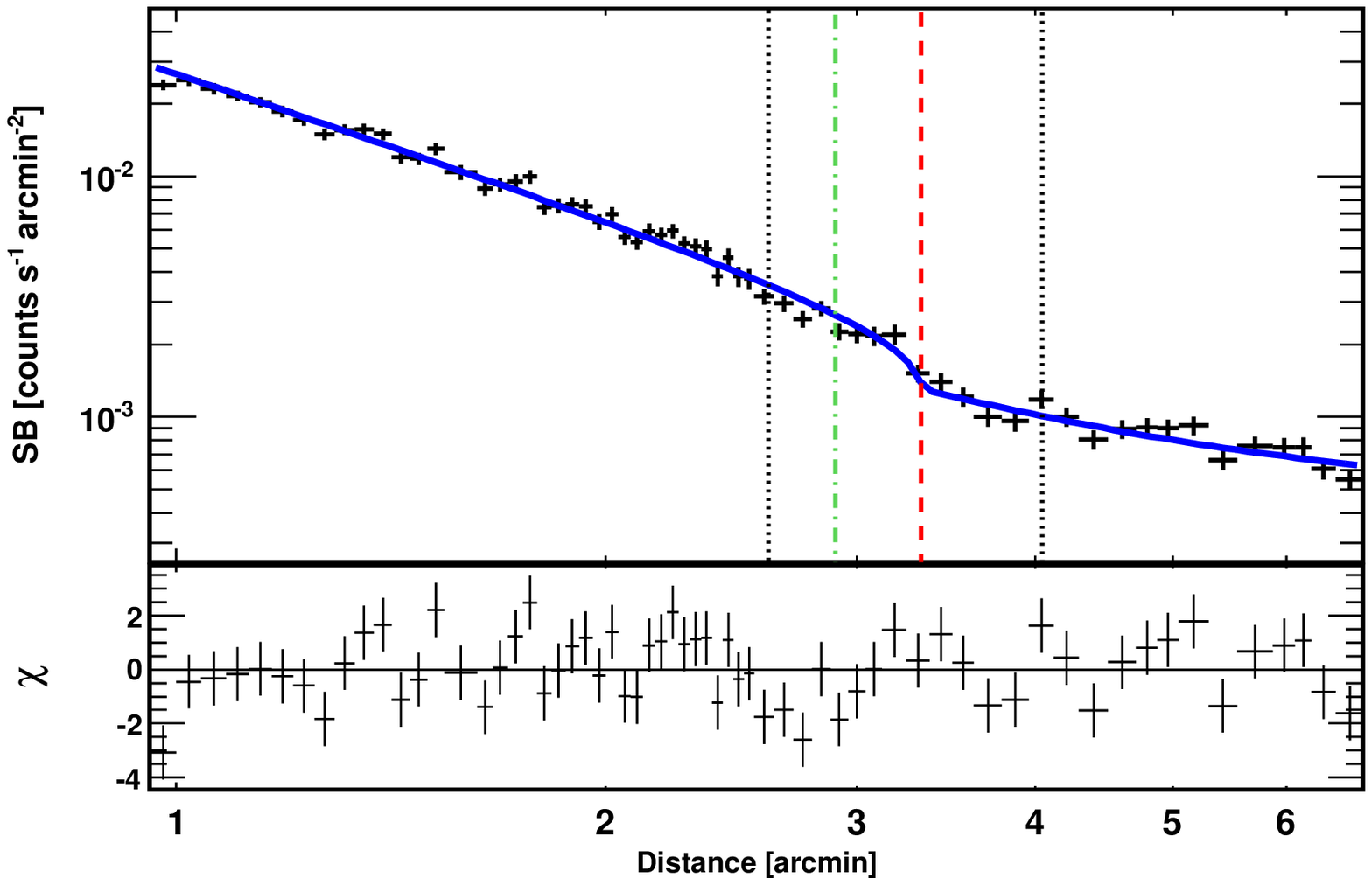}
 \end{center}
 \caption{\emph{Top:} {Instrumental-background-subtracted, exposure-corrected surface brightness profiles across the northern (left) and southern (right) sectors shown in Figure \ref{fig:shock-pies}.} The purple lines show the instrumental background in the two sectors. Each bin has a SNR of at least 10. \emph{Bottom:} Surface brightness profiles around the density discontinuities, with best-fit models. The models assume sphericity and an underlying density profile described by two power-laws with a jump at the shock front. The sky background value was fixed to $4.4\times 10^{-4}$~counts~s$^{-1}$~arcmin$^{-2}$, the average background value calculated in the \xmm\ region shown in Figure \ref{fig:regions}. For the northern profile, black vertical lines (dotted) mark the inner and outer boundaries of the Toothbrush relic, and the green line (dot-dash) marks the outer boundary of the N-NW radio extension. For the southern profile, black vertical lines (dotted) mark the inner and outer boundaries of the small ($\sim 200$ kpc) SE relic, while the green line (dot-dash) marks the southern boundary of the radio halo. In both of the bottom plots, the best-fit shock radii are marked with red lines (dashed). The surface brightness jumps correspond to Mach numbers of $1.7_{-0.42}^{+0.41}$ (north) and $1.5_{-0.086}^{+0.098}$ (south).}
 \label{fig:shock-profiles}
\end{figure*}

\begin{figure}
 \begin{center}
  \includegraphics[width=0.49\textwidth,keepaspectratio=true,clip=true,trim=1.5cm 1.15cm 2.2cm 3cm]{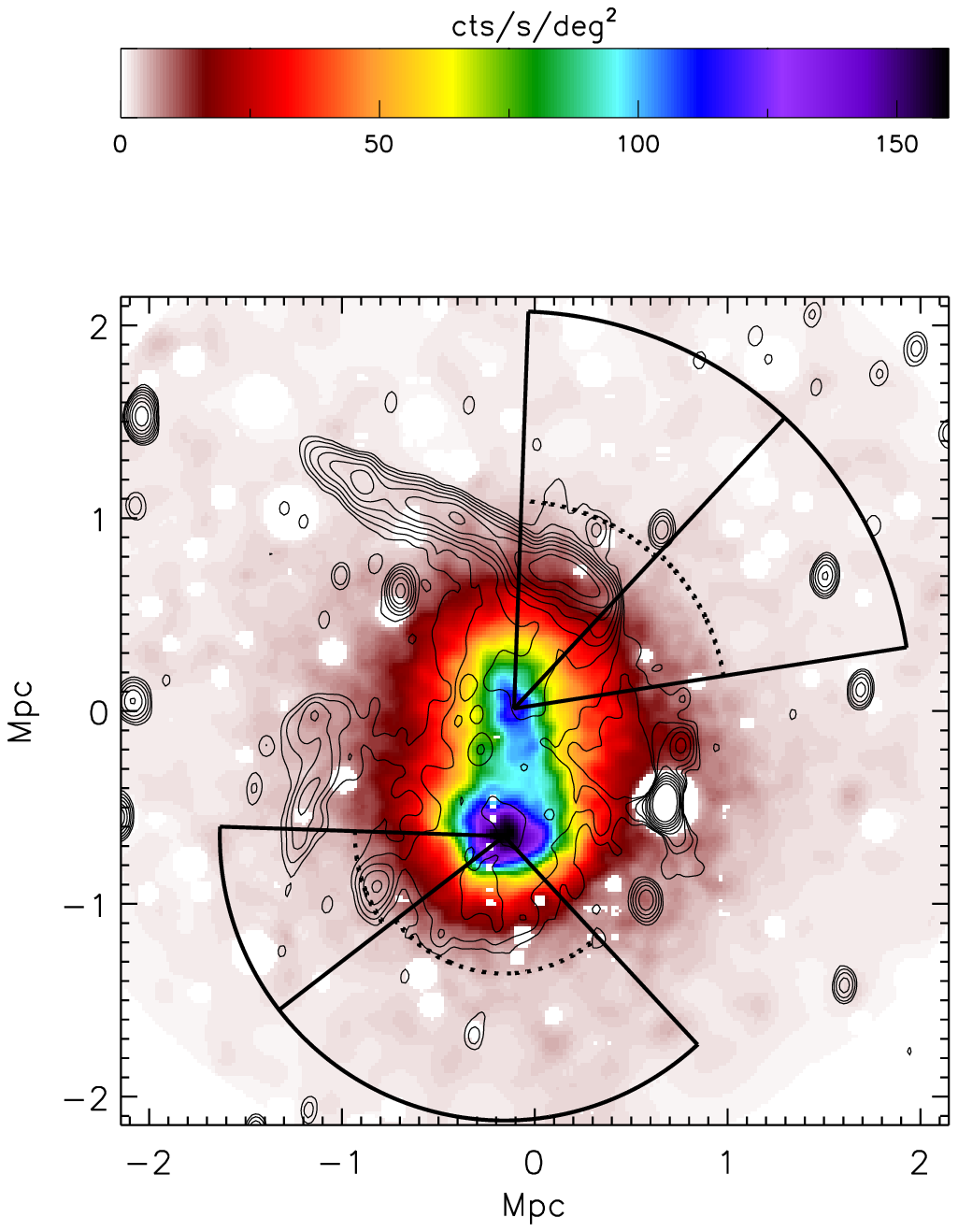}
 \end{center}
 \caption{The adaptively smoothed image shown in Figure \ref{fig:xmmimg-smooth}, with overlaid regions showing the sectors used for extracting the surface brightness profiles in Figures \ref{fig:shock-smallprofiles-n} and \ref{fig:shock-smallprofiles-s}. Dotted lines show the best-fit positions of the density discontinuities. Overlaid are the same radio contours as in Figure \ref{fig:xmmimg-smooth}.}
 \label{fig:shock-pies-small}
\end{figure}

\begin{figure*}
 \begin{center}
  \includegraphics[width=0.49\textwidth,keepaspectratio=true,clip=true]{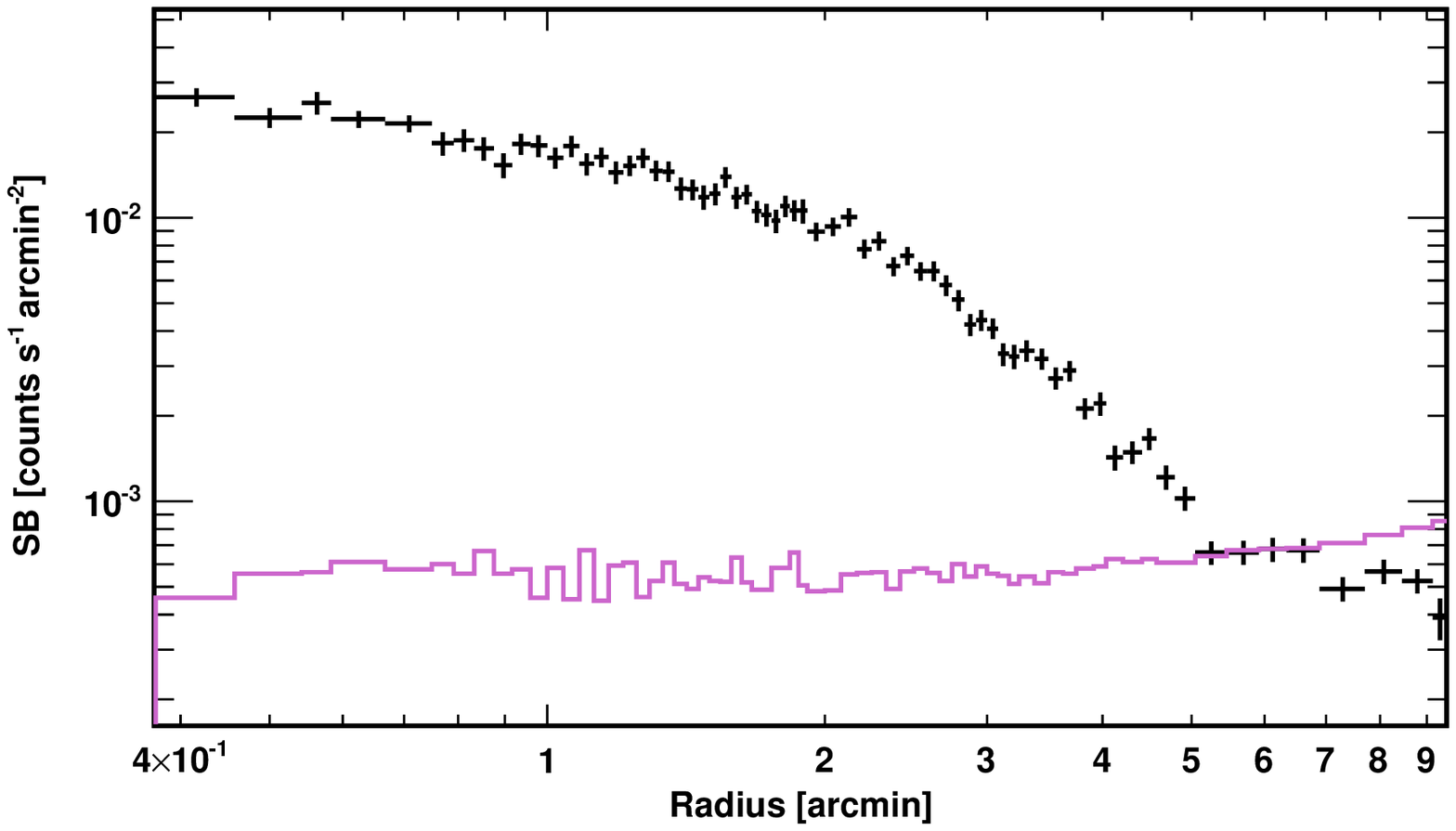}
  \includegraphics[width=0.49\textwidth,keepaspectratio=true,clip=true]{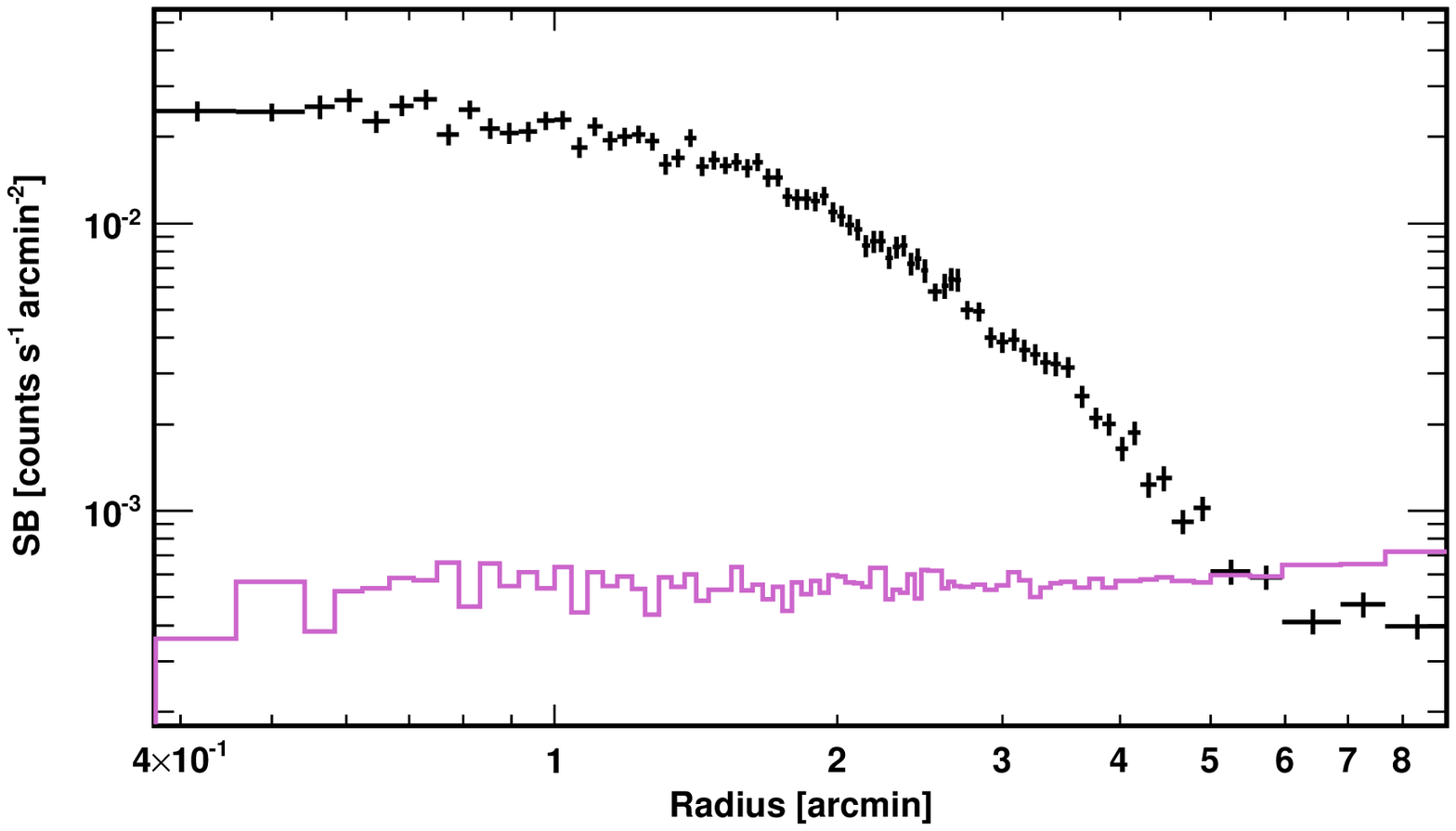}
  \includegraphics[width=0.497\textwidth,keepaspectratio=true,clip=true]{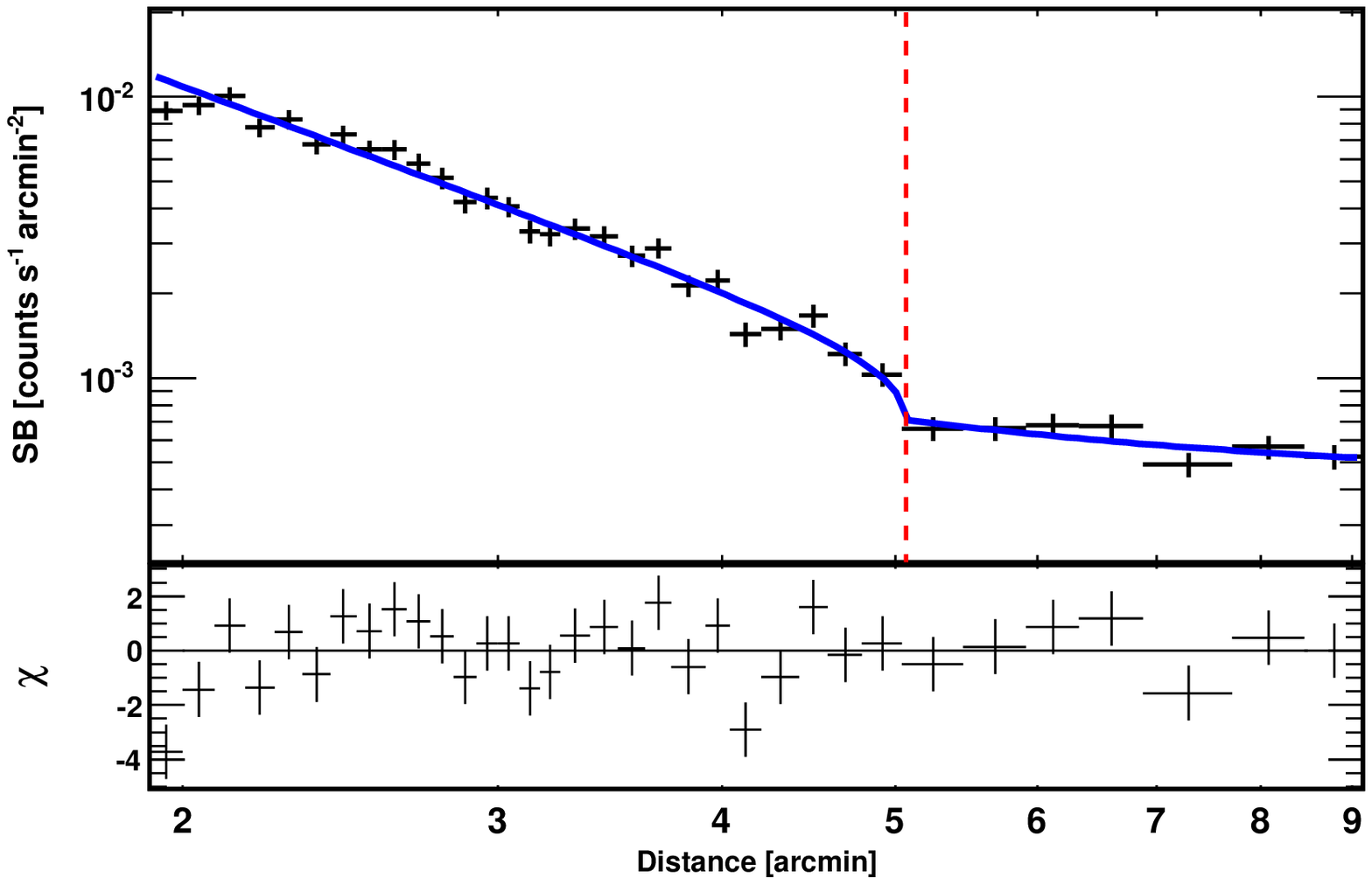}
  \includegraphics[width=0.497\textwidth,keepaspectratio=true,clip=true]{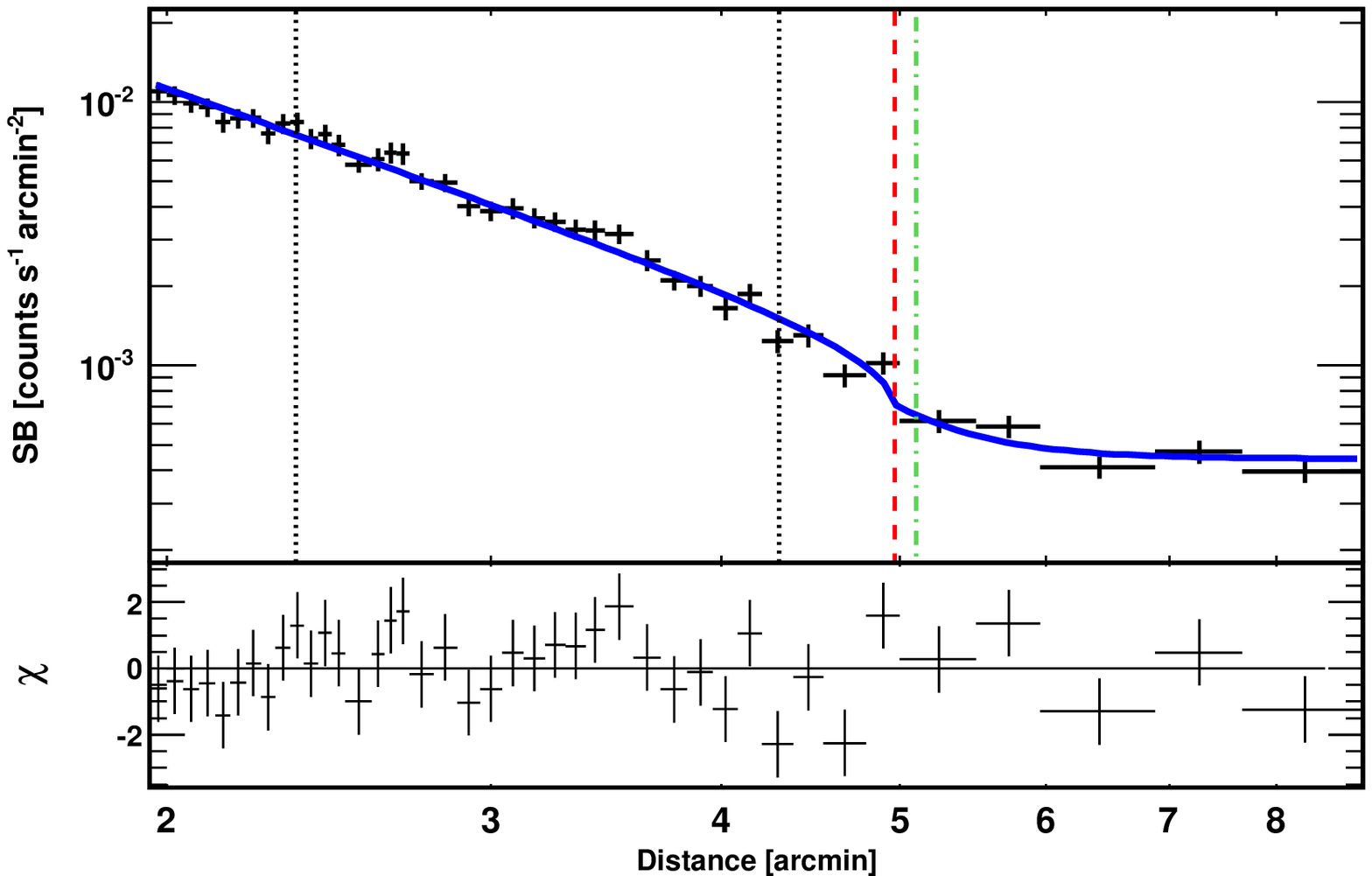}
 \end{center}
 \caption{\emph{Top:} Instrumental-background-subtracted surface brightness profiles across the NW (left) and N (right) sectors shown in Figure \ref{fig:shock-pies-small}. The purple lines show the instrumental background in the two sectors. Each bin has a SNR of at least 10. \emph{Bottom:} Surface brightness profiles around the density discontinuities, with best-fit models. The models are the same as those used for the profiles shown in Figure \ref{fig:shock-profiles}. For the N profile, black vertical lines (dotted) mark the inner and outer boundaries of the Toothbrush relic, and the green line (dot-dash) marks the outer boundary of the N-NW radio extension. In both of the bottom plots, the best-fit shock radii are marked with red lines (dashed). The surface brightness jumps correspond to Mach numbers of $1.9_{-0.42}^{+0.75}$ (NW) and $1.3\pm 0.45$ (N).}
 \label{fig:shock-smallprofiles-n}
\end{figure*}

\begin{figure*}
 \begin{center}
  \includegraphics[width=0.49\textwidth,keepaspectratio=true,clip=true]{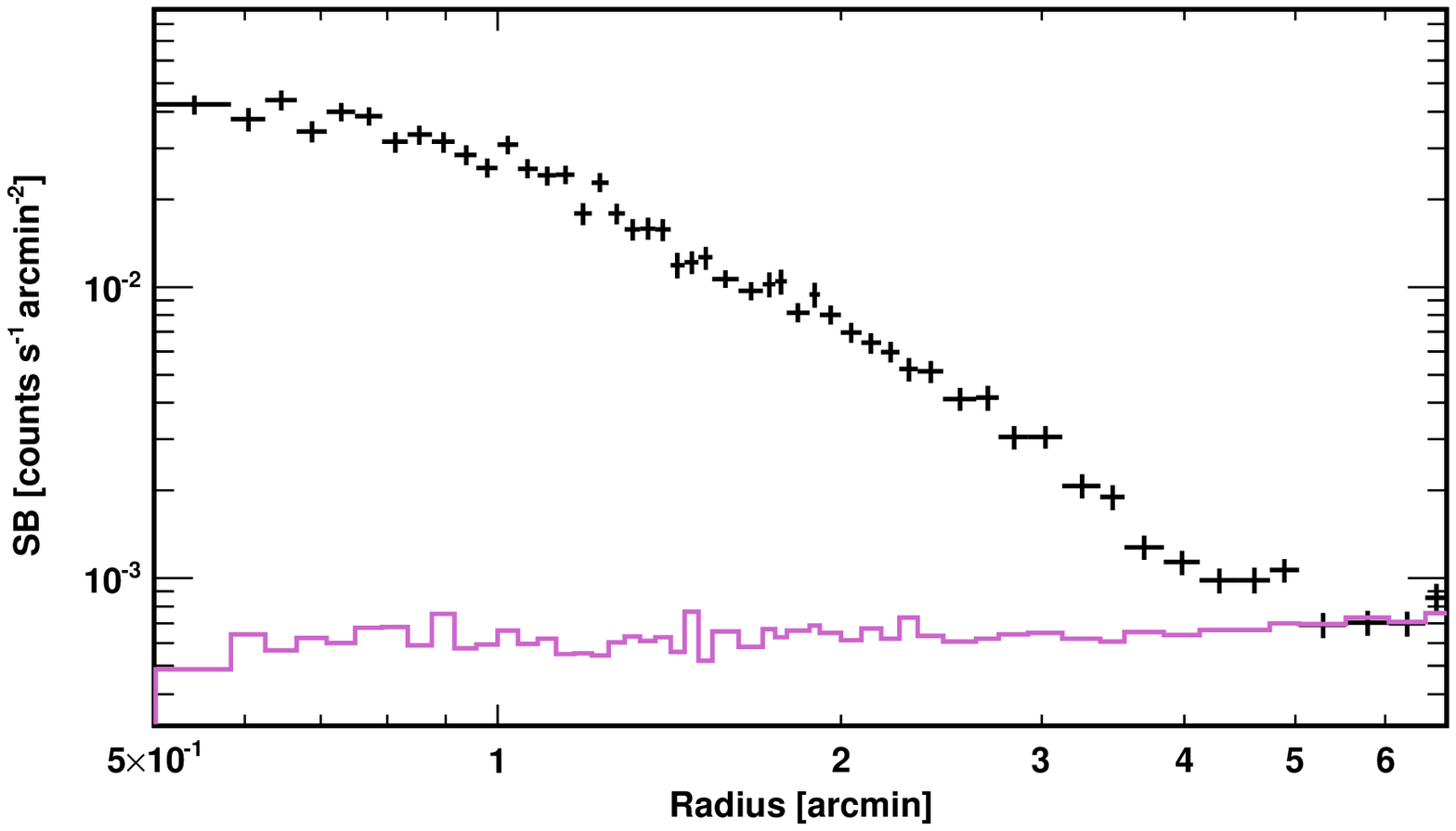}
  \includegraphics[width=0.49\textwidth,keepaspectratio=true,clip=true]{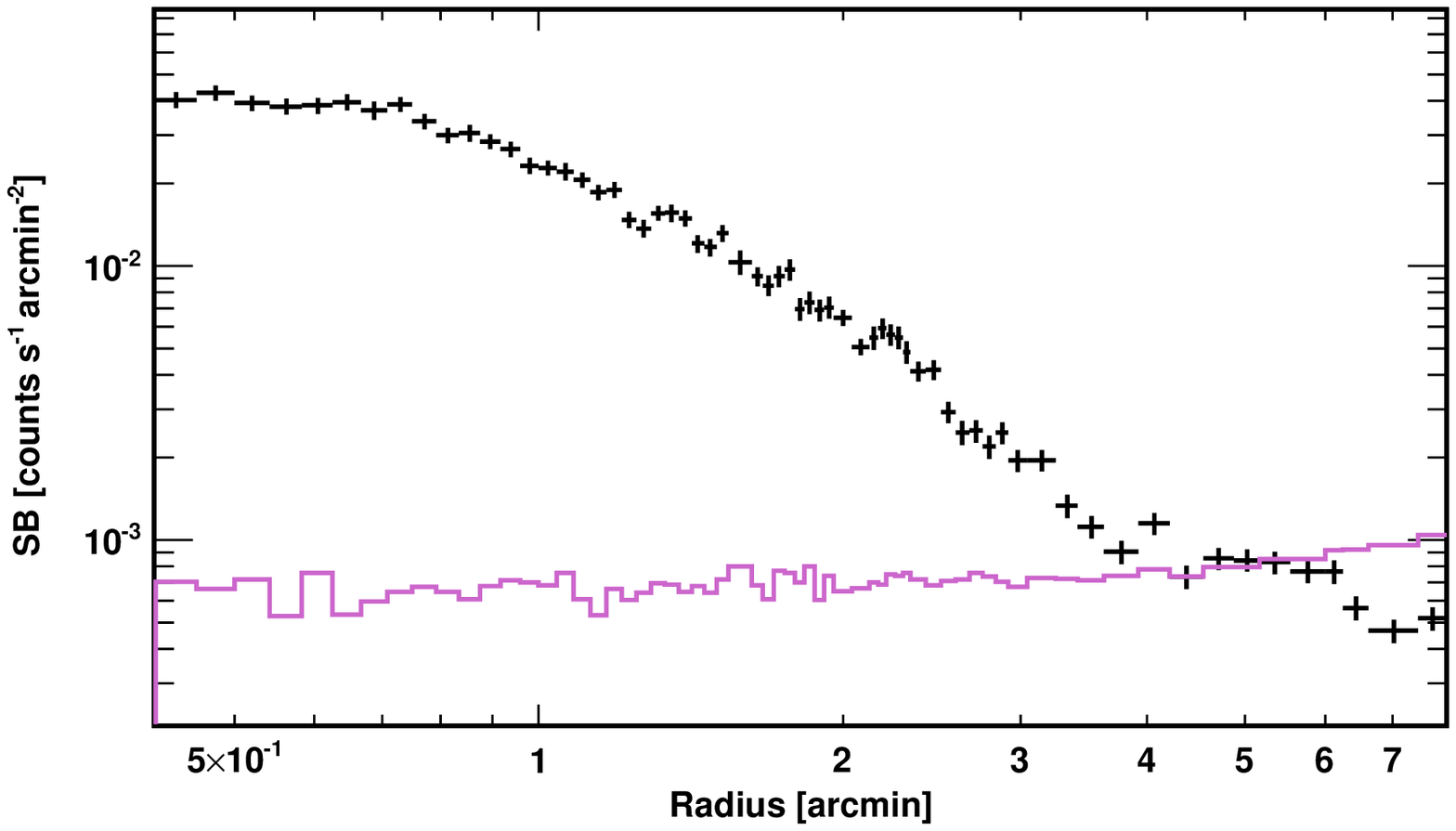}
  \includegraphics[width=0.497\textwidth,keepaspectratio=true,clip=true]{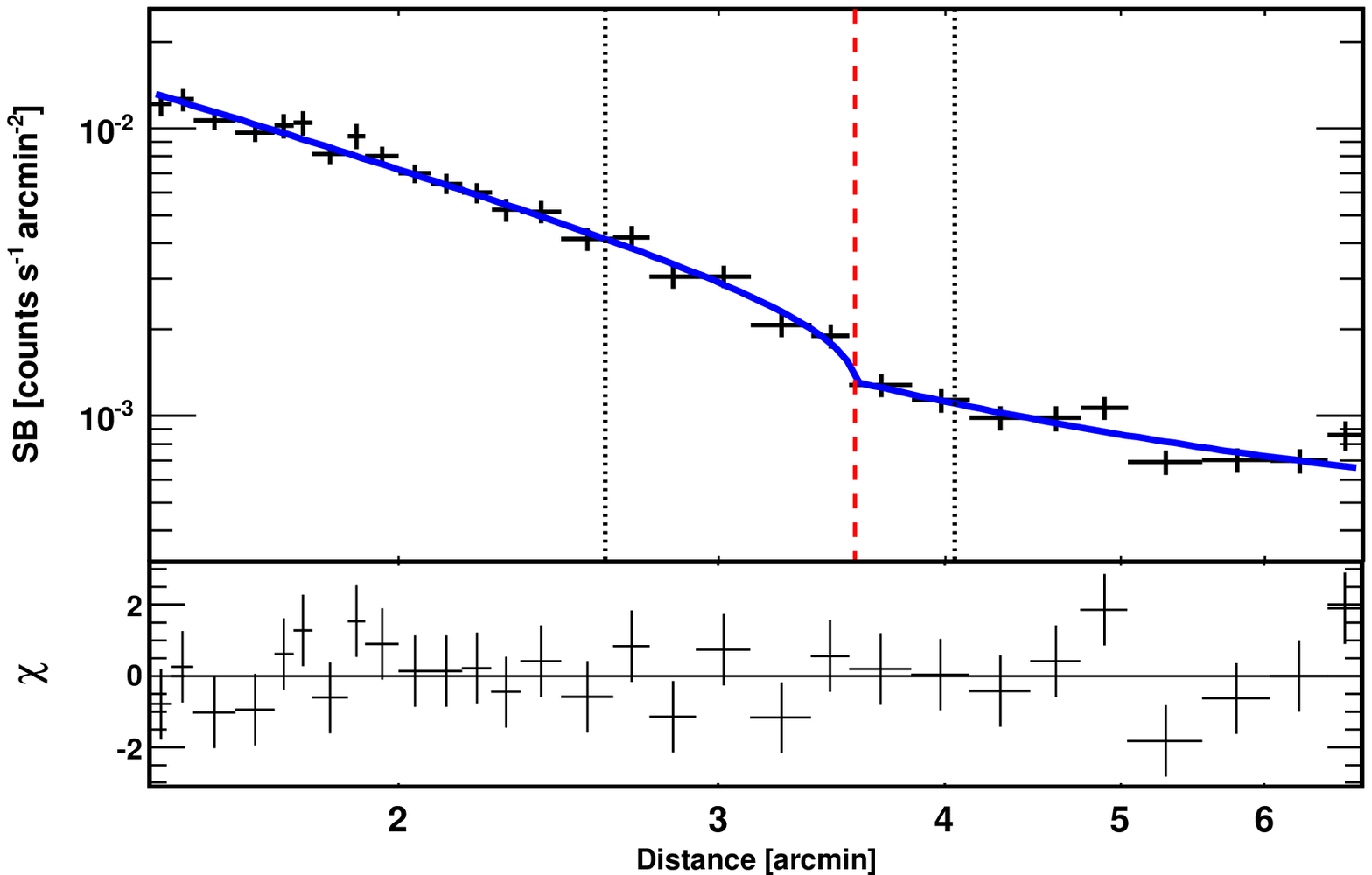}
  \includegraphics[width=0.497\textwidth,keepaspectratio=true,clip=true]{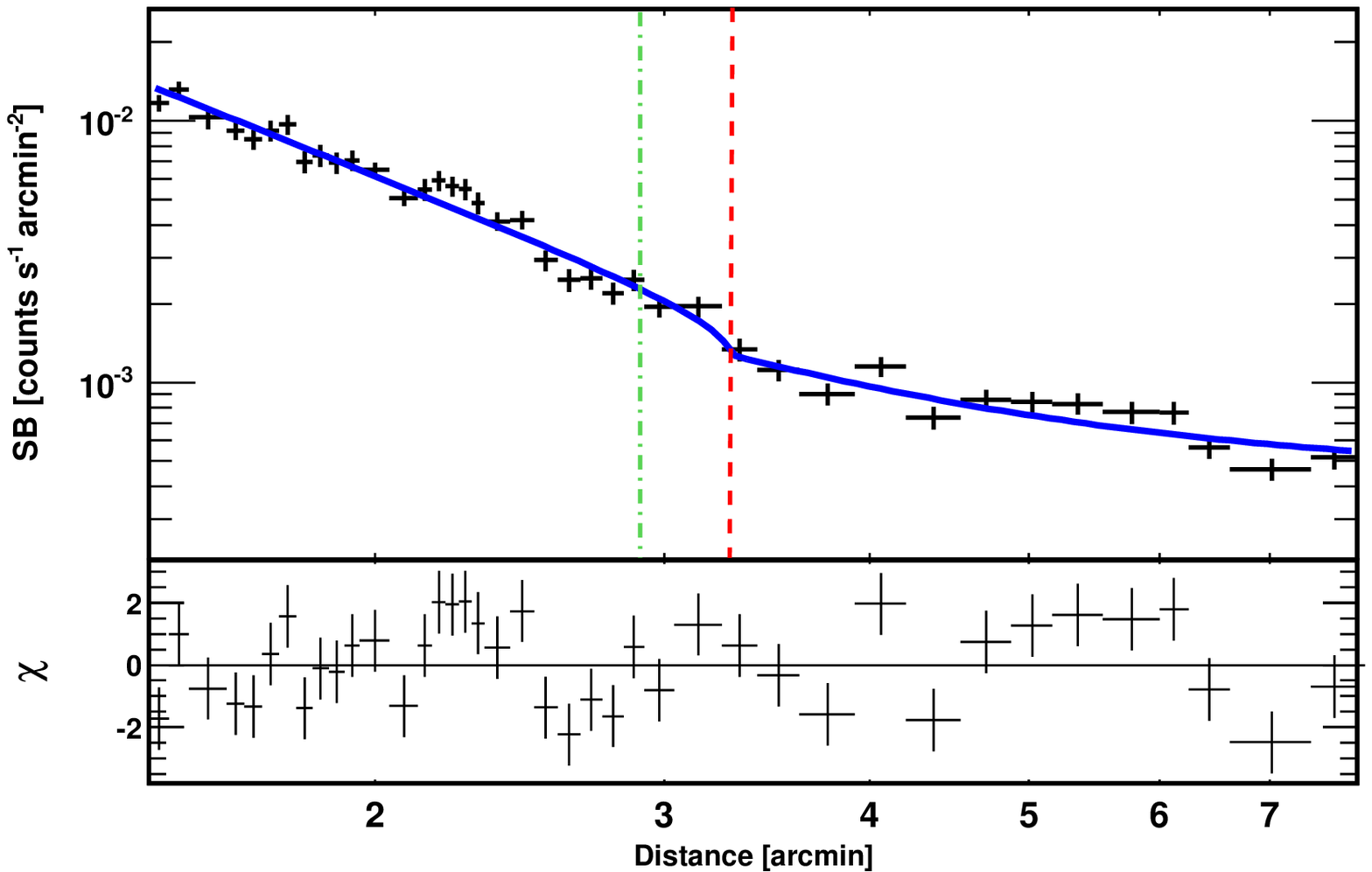}
 \end{center}
 \caption{\emph{Top:} Instrumental-background-subtracted surface brightness profiles across the SE (left) and S (right) sectors shown in Figure \ref{fig:shock-pies-small}. The purple lines show the instrumental background in the two sectors. Each bin has a SNR of at least 10. \emph{Bottom:} Surface brightness profiles around the density discontinuities, with best-fit models. The models are the same as those used for the profiles shown in Figure \ref{fig:shock-profiles}. For the SE profile, black vertical lines (dotted) mark the inner and outer boundaries of the SE relic. In both of the bottom plots, the best-fit shock radii are marked with red lines (dashed). The surface brightness jumps correspond to Mach numbers of $1.5_{-0.13}^{+0.17}$ (SE) and $1.3_{-0.087}^{+0.10}$ (S).}
 \label{fig:shock-smallprofiles-s}
\end{figure*}

\begin{figure}
 \begin{center}
  \includegraphics[width=0.49\textwidth,keepaspectratio=true,clip=true,trim=1.5cm 1.15cm 2.2cm 3cm]{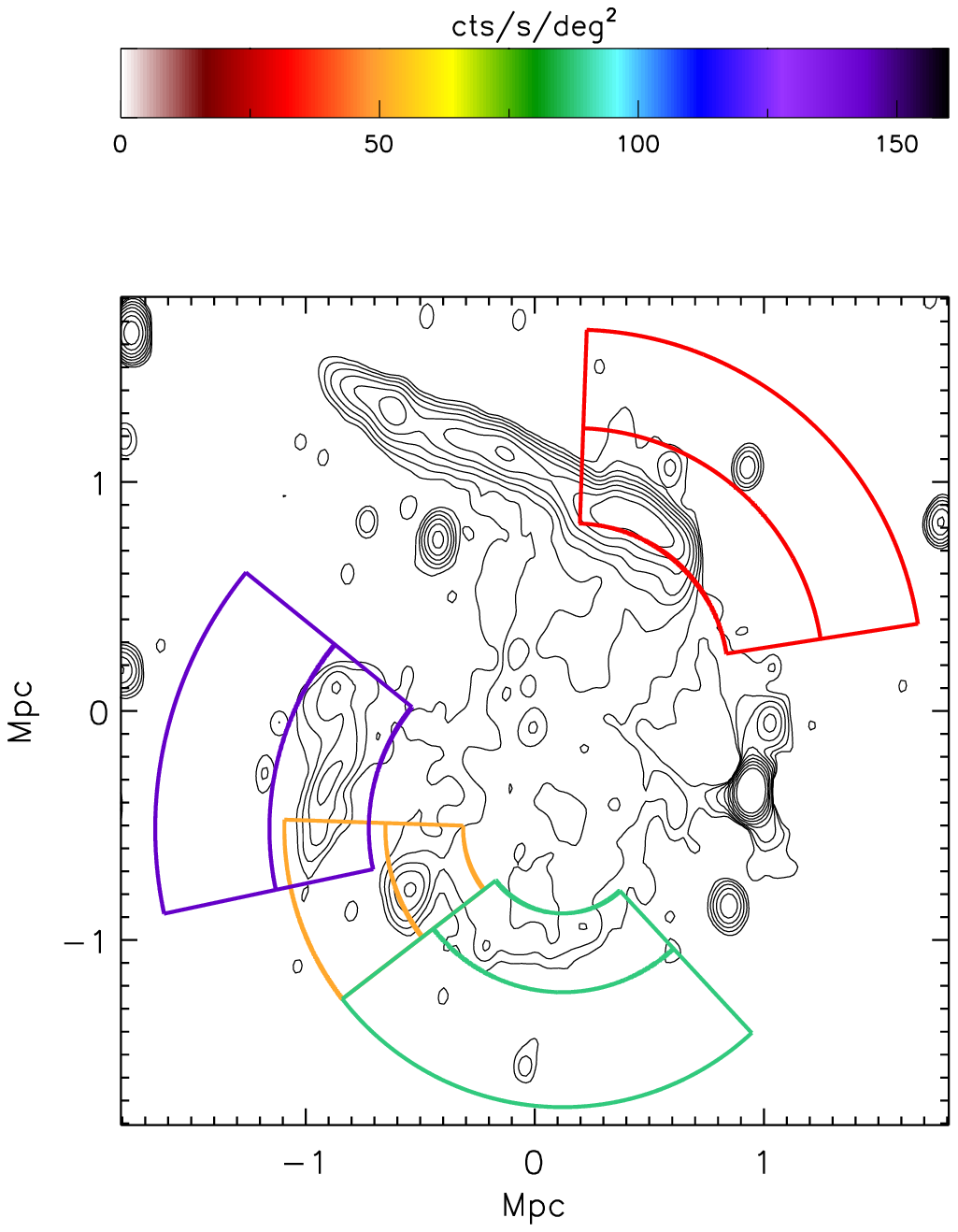}
 \end{center}
 \caption{Regions used for calculating the post- and pre-shock temperatures at the N (red), S (green), SE (orange), and E (purple) shocks. Radio contours are the same as in Figure \ref{fig:xmmimg-smooth}.}
 \label{fig:shock-spectra}
\end{figure}

To examine the surface brightness discontinuities identified in the unsharp-masked image, we extracted EPIC instrumental background-subtracted surface brightness profiles in the sectors shown in Figure \ref{fig:shock-pies}. The annuli in each sector were chosen to have a minimum signal-to-noise radio (SNR) of 10. The resulting profiles are shown in Figure \ref{fig:shock-profiles}. Each profile was fitted with a broken power-law density model in a region around the putative shock front radius, plus a constant describing the sky background. The density models can be summarized as:
\begin{eqnarray}
  n_2 & = & C \, n_0 \left(\frac{r}{r_{\rm shock}}\right)^{\alpha_2} \; , \;\;\;\; {\rm for} \; r \le r_{\rm shock} \nonumber \\
  n_1 & = & n_0 \left(\frac{r}{r_{\rm shock}}\right)^{\alpha_1} \nonumber \; , \;\;\;\; {\rm otherwise.}
\end{eqnarray}
In the equations above, $n$ is the electron number density, $C$ is the shock compression, $r$ is the distance from the centre of the sector, $r_{\rm shock}$ is the shock radius, and $\alpha$ is the power-law index. Subscripts 1 and 2 stand for pre-shock and post-shock parameters, respectively. The emissivity, proportional to the density square, is integrated along the line-of-sight, assuming spherical symmetry, to calculate the surface brightness model. The sky background value was calculated from the region shown in Figure \ref{fig:regions}, and kept fixed in the fit. Fitting was done using PROFFIT v1.1\footnote{http://www.iasf-milano.inaf.it/~eckert/newsite/Proffit.html} \citep{Eckert2011b}. The best-fit models are shown in Figure \ref{fig:shock-profiles}.

As expected from the unsharp-masked image, we find surface brightness discontinuities in each of the two sectors. To the north, the assumed underlying density model indicates shock compression by a factor of $2.0_{-0.49}^{+0.48} $; the fit had $\chi^2/{\rm d.o.f.}=66.2/57$. To the south, the compression factor is $1.7_{-0.11}^{+0.13} $, with $\chi^2/{\rm d.o.f.}= 85.5/57$. The Mach numbers at the northern and southerns discontinuities are therefore relatively low, $1.7_{-0.42}^{+0.41}$ and $1.5_{-0.086}^{+0.098}$, respectively.

Spherical symmetry is one of the main assumptions behind our surface brightness models. However, it is a rather crude approximation, given the clearly disturbed cluster morphology. We can minimize the effects introduced by deviations from spherical symmetry by reducing the opening angle of each sector. Hence, we divided each of the sectors in Figure \ref{fig:shock-pies} into two, and searched for surface brightness jumps in the resulting profiles. The new sectors and profiles are shown in Figures \ref{fig:shock-pies-small}-\ref{fig:shock-smallprofiles-s}.

To the north, across the broadest part of the Toothbrush relic, a very weak density discontinuity is present at the outer boundary of the small N-NW radio extension. This discontinuity corresponds to a Mach number of $1.3\pm 0.45$, therefore consistent with 1. To the NW, the Mach number inferred from the best-fit density jump is higher, $1.9_{-0.42}^{+0.75}$, although the errors are also large. Both the N and NW fits were satisfactory, with $\chi^2/{\rm d.o.f.}=42.9/35$ and $\chi^2/{\rm d.o.f.}=37.9/27$, respectively.

To the south, we divided the larger sector in Figure \ref{fig:shock-pies} such that one profile crosses the SE relic, while one crosses the edge of the radio halo. The SE profile is excellently fitted by our model, with $\chi^2/{\rm d.o.f.}=23.8/23$ and a best-fit density jump of $1.7_{-0.18}^{+0.22}$, corresponding to a Mach number of $1.5_{-0.13}^{+0.17}$. Modelling the S profile indicates a Mach number of $1.3_{-0.087}^{+0.10}$, although the fit is much poorer ($\chi^2/{\rm d.o.f.}=70.7/34$).

We further examined the detected surface brightness discontinuities by extracting spectra in regions tangential to the best-fit shock front radii. These regions are shown in Figure \ref{fig:shock-spectra}. To the N-NW, the difference between the best-fit shock radii in the two sectors shown in Figure \ref{fig:shock-pies} is only $6.1$ arcsec, essentially the spatial resolution limit of \xmm. In order to improve the SNR of the spectra extracted N-NW of the Toothbrush, we have selected two partial annuli that cover the full width of the N-NW sector in Figure \ref{fig:shock-pies}. The boundary between these partial annuli, i.e. the shock radius, is chosen to be the average of the best-fit radii for the narrower N and NW sectors, $r_{\rm shock, N}=5.0$ arcmin. The pre-shock and post-shock spectra were fitted with the same model as the bin spectra, with the sky background parameters fixed to the best-fit values in Table \ref{tab:bkg}. The X-ray column density was fixed to $1.97\times 10^{21}$ cm$^{-2}$, the average value of bins 3 ($N_{\rm H}=1.98\times 10^{21}$ cm$^{-2}$) and 14 ($N_{\rm H}=1.97\times 10^{21}$ cm$^{-2}$), which overlap with the selected post-shock and pre-shock regions. The best-fit post-shock and pre-shock temperatures are $7.0_{-0.46}^{+0.68}$ and $7.1_{-2.4}^{+5.4}$ keV. We performed the same systematic error analysis as described in the previous section. The temperature ranges resulting from $1\sigma$ changes in the sky background and $\pm 5\%$ changes in the QPB are $6.6-7.8$ and $6.5-7.5$ keV, respectively, for the post-shock region, while for the pre-shock region they are $4.3-21.8$ and $4.1-11.1$ keV, respectively. The post-shock temperature is very well-constrained. Although the uncertainties on the pre-shock temperature are significantly larger, they allow us to set a lower limit of approximately $4$ keV. This implies a Mach number $\mathcal{M} \lesssim 1.9$, consistent with the value derived from the surface brightness jump.

As an exercise in caution, we ignored the shock front location determined above, and extracted an additional spectrum from a ``pre-shock'' region tangent to the outer edge of the broadest part of the Toothbrush (Figure \ref{fig:tjump}), rather than to the N-NW radio extension; naively, the outer edge of the relic is the expected shock front location. However, the best-fit temperature in this region was similar to the actual pre-shock temperature, $5.3_{-1.1}^{+1.7}$ keV. Even when possible systematic errors were taken into account, the temperature was still above $3$ keV, ranging between $3.6$ and $9.4$ keV. Hence, the Mach number at the Toothbrush relic, inferred both from the surface brightness jump and from the temperatures on both sides of the shock front, is not larger than approximately $2$.

To the S and SE, we extracted spectra from partial annuli of width $\sim 1.5-2$ arcmin, tangential to the shock front positions shown in Figure \ref{fig:shock-pies-small}. The best-fit post-shock and pre-shock temperatures were $9.9\pm 0.92$ and $8.1_{-2.0}^{+3.8}$ keV to the S, and $10.0_{-1.2}^{+1.3}$ and $6.4_{-1.3}^{+1.8}$ keV to the SE. In the S regions, the poor count statistics do not allow us to detect a possible temperature discontinuity. In the SE regions, the temperature jump corresponds to a Mach number of $1.6_{-0.33}^{+0.44}$, consistent with the Mach number determined from the surface brightness jump. When systematic errors were taken into account for the two SE regions, the best-fit temperatures varied between $9.5-10.5$ keV for the post-shock region, and $5.3-7.9$ keV for the pre-shock region; therefore, the temperature jump is certain, and the Mach number measurement across the SE relic is a robust result.

\begin{figure*}
 \begin{center}
  \includegraphics[width=0.49\textwidth,keepaspectratio=true,clip=true]{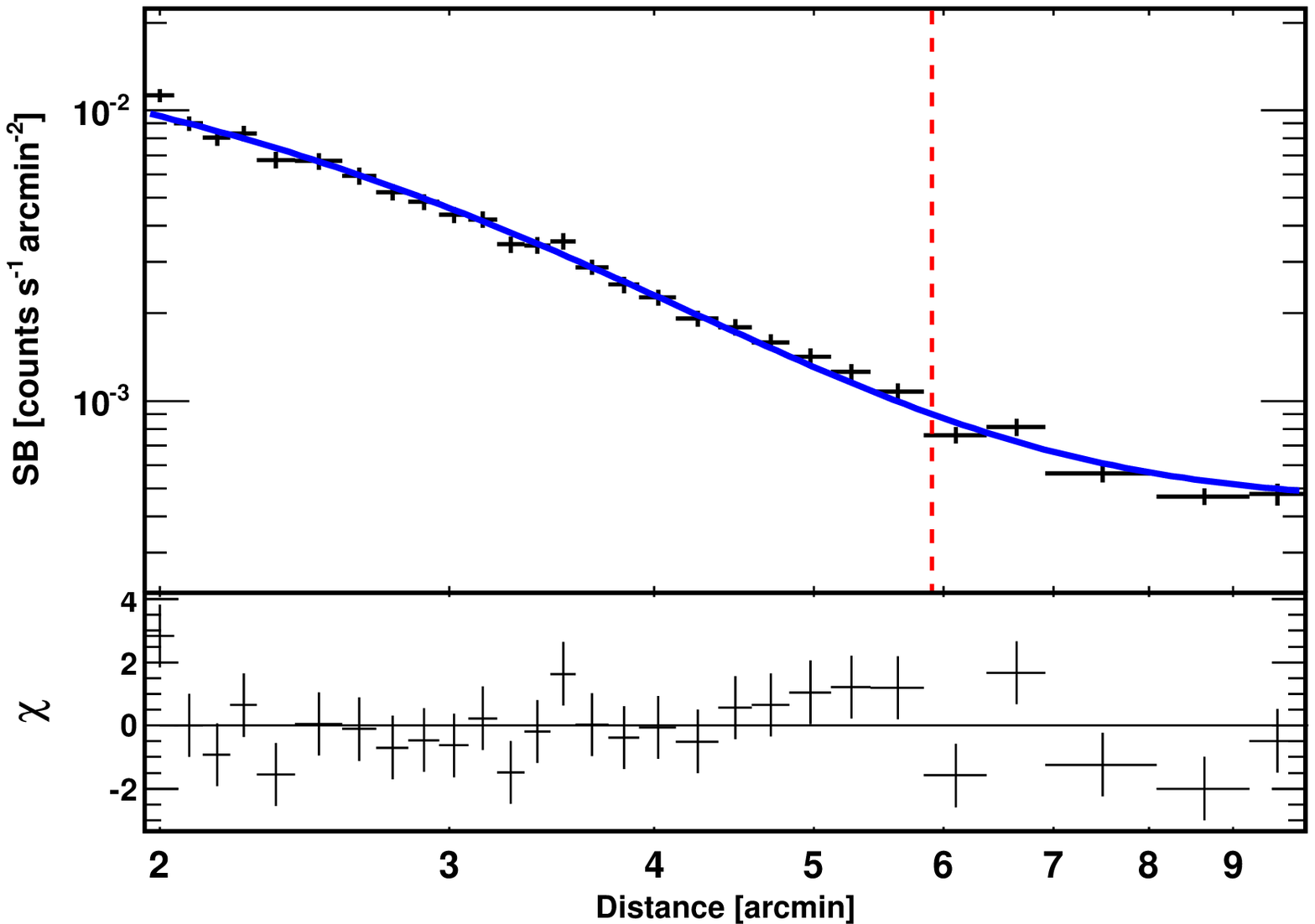}
  \includegraphics[width=0.49\textwidth,keepaspectratio=true,clip=true]{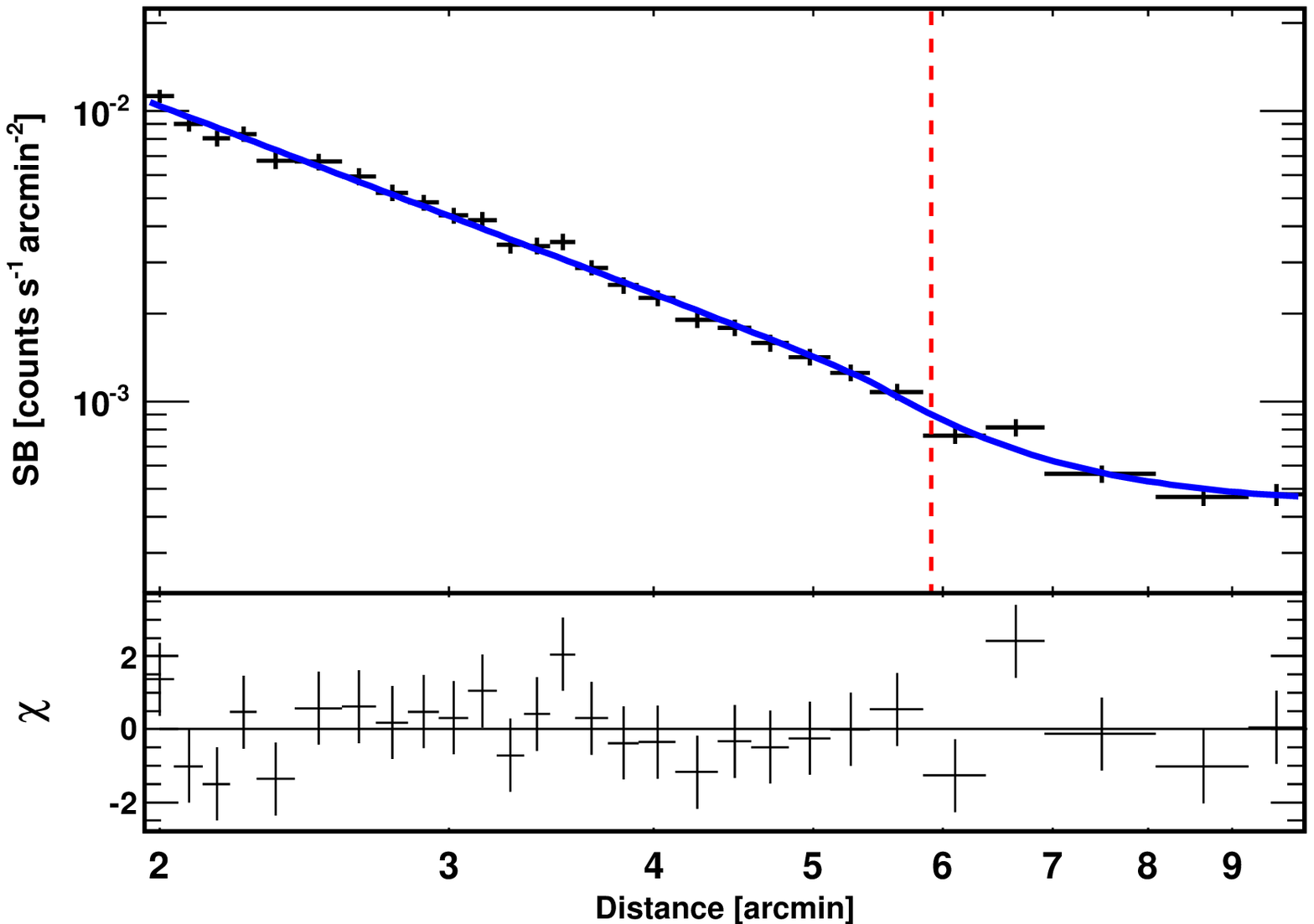}
 \end{center}
 \caption{Instrumental-background-subtracted surface brightness profile in a sector across the eastern relic. The centre of the sector is the same as the centre of the S and SE sectors. Each bin has a SNR of at least 15. Blue lines show the best-fit $\beta$-model (left; $\chi^2/{\rm d.o.f.}=1.4$) and broken-power law model (right; $\chi^2/{\rm d.o.f.}=1.1$). Red dashed lines mark the position of the outer edge of the relic; the same putative shock front radius was used to separate the partial annuli used for measuring the temperature jump across the relic.}
 \label{fig:east}
\end{figure*}

\begin{figure}
 \begin{center}
  \includegraphics[width=0.49\textwidth,keepaspectratio=true,clip=true,trim=1.0cm 1.15cm 2.2cm 3cm]{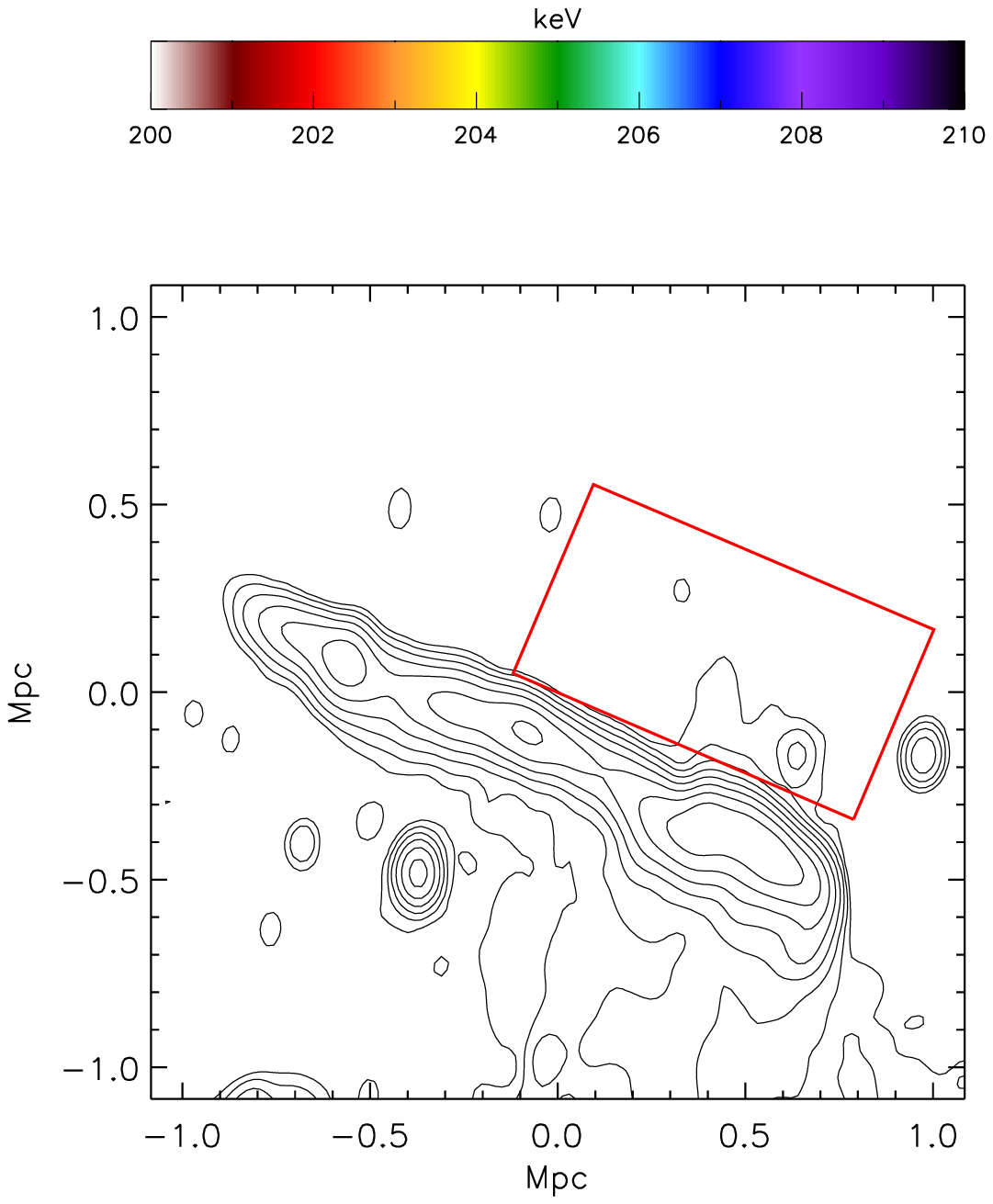}
 \end{center}
 \caption{Additional region used for measuring the temperature ahead of the outer relic edge. This region would be the expected pre-shock region. The best-fit temperature is $5.3_{-1.1}^{+1.7}$. Contours are the same radio contours as in Figure \ref{fig:xmmimg-smooth}.}
 \label{fig:tjump}
\end{figure}

Across the eastern relic, we searched for shock evidence by measuring the temperature in two partial annuli tangent to the outer edge of the relic. The temperature decreases from $8.7_{-1.0}^{+1.5}$ keV in the more central region, to $3.3_{-0.57}^{+0.68}$ keV in the outer region. This jump indicates a shock front of Mach number $2.4_{-0.36}^{+0.46}$. Both the post-shock and pre-shock temperatures are well-constrained, even when systematic errors are taken into account; they range between $7.7-9.8$ and $2.7-4.1$ keV, respectively. However, as can be seen in Figure \ref{fig:east}, there is no surface brightness discontinuity anywhere near the relic. A $\beta$-model is an acceptable fit to the profile, with $\chi^2/{\rm d.o.f.}=35.0/25$. The data is fitted better by a broken power-law model, $\chi^2/{\rm d.o.f.}=25.2/23$, yet although there is a kink in the profile (the power indices are $\alpha_1=1.6_{-0.047}^{+0.042}$ and $\alpha_2=3.2_{-0.35}^{+0.42}$), there is no actual jump (the shock compression is $C=1.0_{-0.11}^{+0.34}$).

\emph{In summary: Surface brightness and temperature discontinuities indicate the presence of three, possibly four, shocks fronts in the ICM, towards N-NW, SE, and E (and perhaps towards S). The N-NW shock is about 1 arcmin ahead of the Toothbrush's outer edge, extends more than half a Mpc beyond the relic's W tip, and has a Mach number significantly below that predicted by the radio spectral index. At the eastern relic, we detect a temperature jump that corresponds to a Mach number of approximately 2.5, but there is no associated surface brightness discontinuity. The surface brightness jump could be masked by projection effects, as was also suggested for the shock front in Abell 665 \citep{Markevitch2001}. The integrated spectral index of the eastern relic is $\alpha=-1.0\pm 0.2$ \citep{vanWeeren2012}, inconsistent with the X-ray-derived Mach number if one assumes an injection spectral index $\alpha_{\rm inj}=\alpha+0.5$, since the $\mathcal{M}(\alpha_{\rm inj})$ function diverges for $\alpha_{\rm inj}=-0.5$. The SE relic has a radio-predicted Mach number of $4.6\pm 1.1$, also significantly above the X-ray result.}

\section{The Toothbrush puzzle}
\label{s:puzzle}

The results presented above pose several challenges to our understanding of radio relics:
\begin{itemize}
  \item Not all relics are associated with surface brightness discontinuities.
  \item The shock front near the Toothbrush is spatially offset from the relic, both in distance and position angle from the cluster centre.
  \item The Mach number of the shock is significantly lower than the Mach number predicted by the injection spectral index.
\end{itemize}
Below we discuss each of these issues.

\subsection{No surface brightness jump}
\label{s:nosxjump}

Although most large-scale shocks are associated with surface brightness discontinuities, there are also exceptions. In Abell 665, \citet{Markevitch2001} identified a shock front in the temperature distribution map, but no corresponding surface brightness discontinuity. At the Coma relic, \citet{Ogrean2012c} and \citet{Akamatsu2013} found a shock of Mach number 2 based on the temperature jump, but no clear surface brightness discontinuity is seen in the surface brightness profile \citep{Ogrean2012c}.

At the eastern relic in the Toothbrush cluster, the temperature jump across the relic's outer edge implies a Mach number of $2.4_{-0.36}^{+0.46}$. Using the Rankine-Hugoniot jump conditions, the Mach number corresponds to a density jump by a factor larger than 2. However, this jump is not reflected in the surface brightness profile; spherical symmetry implies a shock compression consistent with 1, with a possible kink in the underlying density profile. The assumption of spherical symmetry in such a disturbed system is clearly not valid. Therefore, the lack of a surface brightness jump is not equivalent to the lack of a density jump. Even if a density jump was present across the shock, it could potentially be masked by deviations from sphericity and substructure along the line of sight. Their effect on the temperature measurements would be less pronounced. 

\subsection{Relic-shock offset}

As also mentioned in Section \ref{s:intro}, radio relics have only been discovered in merging galaxy clusters, pointing to an association between the two. It is believed that shocks triggered during cluster mergers can accelerate electrons to relativistic energies, and these accelerated electrons gyrate around magnetic field lines, emitting synchrotron radiation at radio frequencies. The acceleration mechanism is believed to be diffusive shock acceleration. Assuming direct acceleration at the shock, the shock front location is expected to be at the outer edge of the relic. As shown in the previous section, this is not the case at the Toothbrush. The shock front position does not coincide with the outer edge of the Toothbrush, with a confidence of approximately $7\sigma$. Even more puzzling, while the relic extends towards the NE, the shock front extends towards the SW, although the two appear to be roughly parallel. There are several possible explanations for the offset:
\begin{itemize}
  \item projection effects;
  \item changes in magnetic field between the relic and the shock front;
  \item a pre-existing population of CR electrons;
  \item a non-detected density discontinuity at the outer edge of the Toothbrush.
\end{itemize}

Projection effects are unlikely though to explain the 1-arcmin offset between the shock and the relic. If relics are created by direct acceleration at the shock, the shock is co-spatial with the outer edge of the relic. Assuming the shock and the relic are spherical caps, it would be possible to detect the shock front on the projected surface of a broad relic. However, it would be impossible to project the detected shock front ahead of the relic.

From equipartition arguments, the magnetic field at the Toothbrush relic is $7-9$ $\mu$G \citep{vanWeeren2012}. If the magnetic field strength dropped between the observed locations of the shock front and the Toothbrush, then synchrotron emission might not be observable in that region, creating an offset between the shock and the relic. However, magnetic fields are compressed at the shock \citep[and possibly also amplified by downstream turbulence, as seen at higher Mach number shocks; e.g.,][]{Inoue2009}, so it is unclear how the magnetic field could drop suddenly in front of the Toothbrush.

Another possibility is that the relic is the result of re-acceleration of a pre-existing CR electron population. One problem with the simplest DSA model is that the particle acceleration efficiency is too low to explain the radio brightness of relics \citep[e.g., ][]{Kang2007,Kang2012}. Consequently, the model has been challenged by numerical simulations showing that a pre-existing population of CR electrons makes it easier to naturally explain the observed radio properties \citep[e.g., ][]{Kang2012,Pinzke2013}. CRs are expected to be formed by, e.g., large scale structure formation shocks \citep[e.g.,][]{Ryu2003}, turbulence \citep[e.g.,][]{Brunetti2001}, hadronic collisions of CR protons with thermal protons in the ICM \citep[e.g.,][]{Dolag2000}, active galactic nuclei \citep[e.g.,][]{George2008}, and supernovae \citep[e.g.,][]{Blasi2011}, and they will remain trapped in the cluster's potential well. If a pre-existing CR population was present only at the location where the relic is seen, then as the shock observed to the N-NW crossed this region, it would have re-accelerated the CRs to energies where they become visible in the radio. On the other hand, beyond the location of the CR population, the efficiency of particle acceleration from the thermal pool into the non-thermal CR population would be low, so no radio emission would be detected.

An unusual situation would be the presence of a density edge at the relic, in addition to the shock detected further out from the cluster centre. This edge could either be a separate shock, or we could be looking at a complex shock surface for which a surface brightness discontinuity is seen only offset from the relic's outer edge. Figures \ref{fig:shock-profiles} and \ref{fig:shock-smallprofiles-n} do not show a surface brightness discontinuity at the relic's outer edge. Yet, as we discuss in Section \ref{s:nosxjump}, multiple effects and assumptions can mask a potential density discontinuity.

\subsection{Mach number discrepancy}

In the linear test-particle regime, DSA predicts that the injection spectral index at the shock front, $\alpha$ ($\mathcal{F}\propto \nu^{\alpha}$), is related to the shock Mach number, $\mathcal{M}$, via 
\begin{eqnarray}
\label{eq:machnumber}
\mathcal{M}^2=(2\alpha-3) / (2\alpha+1) 
\end{eqnarray}
for plane-parallel shocks. If particles are accelerated directly from the thermal pool and the assumptions of the test-particle regime hold, then the radio-derived Mach number should be consistent with the Mach number derived from X-ray data using the Rankine-Hugoniot jump conditions. Indeed, this is the case for all the other merger shocks confirmed at radio relics. In Abell 3667, the spectral index near the outer edge of the radio relic is approximately $-0.7$ \citep{Roettgering1997}, corresponding to a Mach number of approximately $3.3$; the X-ray-derived Mach number at the relic was found to be $2.4\pm 0.77$ \citep{finoguenov10} from the temperature jump across the shock front (we cite the Mach number indicated by the temperature jump, as temperature is less affected by projection effects than density). In Abell 754, the integrated radio spectral index is between $-2.0$ (for the frequency range $0.3-1.4$ GHz) and $-1.8$ (for the frequency range $0.3-0.7$ GHz); if the injection spectral index is $\alpha_{\rm inj}=\alpha+0.5$ \citep{Miniati2002}, the spectral indices at these frequencies correspond to a Mach number of $1.7-1.9$, consistent with the Mach number of $2.1_{-0.6}^{+0.7}$ determined from the temperature jump \citep{Macario2011}. In CIZA J2242.8+5301, the so-called ``Sausage'' relic has a radio-predicted Mach number of $4.6\pm 1.1$ \citep{vanWeeren2010}, consistent with the Mach number of $3.2\pm 0.52$ determined from Suzaku observations near the relic \citep{Akamatsu2011}. The western relic in Abell 3376 has a spectral index that corresponds to a Mach number of $2.2\pm 0.4$ \citep{Kale2012}, which is consistent with the X-ray-derived Mach number of $2.9\pm 0.6$ \citep{Akamatsu2011}.  In Abell 521, the radio spectral index predicts a Mach number of approximately 2.3 \citep{Giacintucci2008}, consistent with the Mach number of $3.4_{-1.9}^{+3.7}$ determined from X-ray \citep{Bourdin2013}. 

The injection spectral index at the Toothbrush relic is between $-0.7$ and $-0.6$, corresponding to $\mathcal{M}=3.3-4.6$. The Mach number derived both from the density and the temperature discontinuities across the shock front is no larger than 2, significantly lower than the simplest DSA predictions. A possible explanation is that any shock front is comprised of a distribution of Mach numbers \citep[e.g.,][]{Skillman2013}. Synchrotron emission is more sensitive to high Mach numbers \citep{HoeftBrueggen2007}, so shock regions of high Mach number can introduce an upward bias in the overall Mach number derived from radio observations.

Moreover, \citet{Kang2012} have shown that the Sausage relic, which has a radio-predicted Mach number of $4.6\pm 1.1$ \citep{vanWeeren2010}, can be also explained by a shock front with Mach number of about 2, if a population of pre-existing CR electrons is introduced in the model. While such a low Mach number at the Sausage relic has been shown not to be consistent with observations \citep{Akamatsu2011}, the conclusions of \citet{Kang2012} suggest that one possible explanation for the Mach number discrepancy at the Toothbrush relic is that the shock re-accelerated a pre-existing population of CR electrons.

Alternative explanations also exist. Projection effects, for example, could smooth out both the density and the temperature discontinuities. Furthermore, the Mach number range of $3.3-4.6$ was derived from the radio spectral index assuming a plane-parallel shock. For oblique shocks, the simple equation relating the Mach number and the spectral index (Eq. \ref{eq:machnumber}) over-predicts the strength of the shock \citep[e.g.,][]{Kirk1989}.

\emph{The shock front detected at the Toothbrush relic challenges our current understanding of radio relics. It is spatially offset from the radio relic, and its Mach number is significantly below that predicted from the spectral index at the front of the relic assuming DSA in the test particle regime. It appears challenging to explain the observed shock characteristics using the simplest DSA model, in which CR electrons at the relic are accelerated directly from the thermal pool. However, the observations alone cannot exclude alternative explanations such as projection effects or complex magnetic fields, and numerical simulations would be required to test the different hypothesis.}

\section{Summary}
\label{s:conclusions}

The Toothbrush cluster ($z=0.225$) hosts three radio relics and a faint radio halo. The northern relic, the Toothbrush, is the most spectacular of the three relics due to its unusual linear shape, which might have been caused by a triple merger event. The spectral index at the front of the relic suggests a Mach number between 3.3 and 4.6. Here, we analyzed a deep \xmm\ observation of the cluster, identified the shock fronts present in the ICM, and compared their properties with those predicted by radio observations. Below is a summary of our main results:

\begin{itemize}
  \item We find clear evidence both from imaging and spectral results that the Toothbrush cluster is a merging system. The cores of the merging clusters survived the merger, and are now observed at a separation of 650 kpc, connected by a bridge of dense plasma.
  \item The temperature decreases from W to E along the Toothbrush.
  \item Surface brightness discontinuities are present N-NW of the subcluster's core, and S-SE of the more massive core.
  \item We find evidence for three, possibly four, weak shock fronts. They are found towards N, E, and SE (and possibly S), near the positions of previously discovered radio relics, and have Mach numbers lower than about $2.5$. All three Mach numbers are smaller than the Mach numbers predicted from the radio spectral indices under the assumptions of direct CR injection and DSA in the linear test particle regime. The shocks in the Toothbrush cluster are the only ones with such clear differences between the radio-predicted and X-ray-derived Mach numbers, and a larger relic sample with very high quality radio and X-ray data would be required for a statistical analysis of such occurences. With the advent of new radio arrays such as LOFAR and SKA, more relics with very high Mach numbers predicted by their spectral indices could prove instead to be associated with much weaker shocks.
  \item The N-NW shock, which was expected to trace the outer edge of the Toothbrush relic, is unlike any of the other shocks discovered so far at radio relics. It is offset by about 1 arcmin NW from the edge of the Toothbrush, and extends by about 700 kpc towards SW, beyond the W tip of the relic. Numerical simulations will be required to explain the shock's unusual properties in relation to the relic.
\end{itemize}

\section*{Acknowledgments}

GAO thanks Franco Vazza, Annalisa Bonafede, and Elke Roediger for helpful discussions. MB and MH acknowledge support by the research group FOR 1254 funded by the Deutsche Forschungsgemeinschaft (DFG). RJvW acknowledges support provided by NASA through Einstein Postdoctoral Fellowship grant number PF2-130104 awarded by the Chandra X-ray Center, which is operated by the Smithsonian Astrophysical Observatory for NASA under contract NAS8-03060. This research is based on data from observations obtained with \xmm, an ESA science mission with instruments and contributions directly funded by ESA Member States and the USA (NASA).

\bibliographystyle{mn2e}
\bibliography{bibliography}

\begin{thebibliography}{}

\bibitem[\protect\citeauthoryear{{Akamatsu}, {Inoue}, {Sato}, {Matsushita},
  {Ishisaki} \& {Sarazin}}{{Akamatsu} et~al.}{2013}]{Akamatsu2013}
{Akamatsu} H.,  {Inoue} S.,  {Sato} T.,  {Matsushita} K.,  {Ishisaki} Y.,
  {Sarazin} C.~L.,  2013, ArXiv e-prints

\bibitem[\protect\citeauthoryear{{Akamatsu} \& {Kawahara}}{{Akamatsu} \&
  {Kawahara}}{2011}]{Akamatsu2011}
{Akamatsu} H.,  {Kawahara} H.,  2011, ArXiv e-prints

\bibitem[\protect\citeauthoryear{{Anders} \& {Grevesse}}{{Anders} \&
  {Grevesse}}{1989}]{angr1989}
{Anders} E.,  {Grevesse} N.,  1989, \gca, 53, 197

\bibitem[\protect\citeauthoryear{{Blandford} \& {Eichler}}{{Blandford} \&
  {Eichler}}{1987}]{Blandford1987}
{Blandford} R.,  {Eichler} D.,  1987, \physrep, 154, 1

\bibitem[\protect\citeauthoryear{{Blasi}}{{Blasi}}{2011}]{Blasi2011}
{Blasi} P.,  2011, in {Giani} S.,  {Leroy} C.,   {Rancoita} P.~G.,  eds, Cosmic
  Rays for Particle and Astroparticle Physics. pp 493--506

\bibitem[\protect\citeauthoryear{{Bourdin}, {Mazzotta}, {Markevitch},
  {Giacintucci} \& {Brunetti}}{{Bourdin} et~al.}{2013}]{Bourdin2013}
{Bourdin} H.,  {Mazzotta} P.,  {Markevitch} M.,  {Giacintucci} S.,
  {Brunetti} G.,  2013, \apj, 764, 82

\bibitem[\protect\citeauthoryear{{Br{\"u}ggen}, {van Weeren} \&
  {R{\"o}ttgering}}{{Br{\"u}ggen} et~al.}{2012}]{Brueggen2012}
{Br{\"u}ggen} M.,  {van Weeren} R.~J.,    {R{\"o}ttgering} H.~J.~A.,  2012,
  \mnras, 425, L76

\bibitem[\protect\citeauthoryear{{Brunetti}, {Setti}, {Feretti} \&
  {Giovannini}}{{Brunetti} et~al.}{2001}]{Brunetti2001}
{Brunetti} G.,  {Setti} G.,  {Feretti} L.,    {Giovannini} G.,  2001, \mnras,
  320, 365

\bibitem[\protect\citeauthoryear{{De Luca} \& {Molendi}}{{De Luca} \&
  {Molendi}}{2004}]{delucamolendi2004}
{De Luca} A.,  {Molendi} S.,  2004, \aap, 419, 837

\bibitem[\protect\citeauthoryear{{Diehl} \& {Statler}}{{Diehl} \&
  {Statler}}{2006}]{DiehlStatler2006}
{Diehl} S.,  {Statler} T.~S.,  2006, \mnras, 368, 497

\bibitem[\protect\citeauthoryear{{Dolag} \& {En{\ss}lin}}{{Dolag} \&
  {En{\ss}lin}}{2000}]{Dolag2000}
{Dolag} K.,  {En{\ss}lin} T.~A.,  2000, \aap, 362, 151

\bibitem[\protect\citeauthoryear{{Eckert}, {Molendi} \& {Paltani}}{{Eckert}
  et~al.}{2011}]{Eckert2011b}
{Eckert} D.,  {Molendi} S.,    {Paltani} S.,  2011, \aap, 526, A79

\bibitem[\protect\citeauthoryear{{Feretti}, {Giovannini}, {Govoni} \&
  {Murgia}}{{Feretti} et~al.}{2012}]{Feretti2012}
{Feretti} L.,  {Giovannini} G.,  {Govoni} F.,    {Murgia} M.,  2012, \aapr, 20,
  54

\bibitem[\protect\citeauthoryear{{Finoguenov}, {Sarazin}, {Nakazawa}, {Wik} \&
  {Clarke}}{{Finoguenov} et~al.}{2010}]{finoguenov10}
{Finoguenov} A.,  {Sarazin} C.~L.,  {Nakazawa} K.,  {Wik} D.~R.,    {Clarke}
  T.~E.,  2010, \apj, 715, 1143

\bibitem[\protect\citeauthoryear{{George}, {Fabian}, {Baumgartner}, {Mushotzky}
  \& {Tueller}}{{George} et~al.}{2008}]{George2008}
{George} M.~R.,  {Fabian} A.~C.,  {Baumgartner} W.~H.,  {Mushotzky} R.~F.,
  {Tueller} J.,  2008, \mnras, 388, L59

\bibitem[\protect\citeauthoryear{{Giacintucci} et~al.,}{{Giacintucci}
  et~al.}{2008}]{Giacintucci2008}
{Giacintucci} S.  et~al., 2008, \aap, 486, 347

\bibitem[\protect\citeauthoryear{{Hoeft} \& {Br{\"u}ggen}}{{Hoeft} \&
  {Br{\"u}ggen}}{2007}]{HoeftBrueggen2007}
{Hoeft} M.,  {Br{\"u}ggen} M.,  2007, \mnras, 375, 77

\bibitem[\protect\citeauthoryear{{Inoue}, {Yamazaki} \& {Inutsuka}}{{Inoue}
  et~al.}{2009}]{Inoue2009}
{Inoue} T.,  {Yamazaki} R.,    {Inutsuka} S.-i.,  2009, \apj, 695, 825

\bibitem[\protect\citeauthoryear{{Kalberla}, {Burton}, {Hartmann}, {Arnal},
  {Bajaja}, {Morras} \& {P{\"o}ppel}}{{Kalberla} et~al.}{2005}]{Kalberla2005}
{Kalberla} P.~M.~W.,  {Burton} W.~B.,  {Hartmann} D.,  {Arnal} E.~M.,  {Bajaja}
  E.,  {Morras} R.,    {P{\"o}ppel} W.~G.~L.,  2005, \aap, 440, 775

\bibitem[\protect\citeauthoryear{{Kale}, {Dwarakanath}, {Bagchi} \&
  {Paul}}{{Kale} et~al.}{2012}]{Kale2012}
{Kale} R.,  {Dwarakanath} K.~S.,  {Bagchi} J.,    {Paul} S.,  2012, ArXiv
  e-prints

\bibitem[\protect\citeauthoryear{{Kang}, {Ryu}, {Cen} \& {Ostriker}}{{Kang}
  et~al.}{2007}]{Kang2007}
{Kang} H.,  {Ryu} D.,  {Cen} R.,    {Ostriker} J.~P.,  2007, \apj, 669, 729

\bibitem[\protect\citeauthoryear{{Kang}, {Ryu} \& {Jones}}{{Kang}
  et~al.}{2012}]{Kang2012}
{Kang} H.,  {Ryu} D.,    {Jones} T.~W.,  2012, \apj, 756, 97

\bibitem[\protect\citeauthoryear{{Kirk} \& {Heavens}}{{Kirk} \&
  {Heavens}}{1989}]{Kirk1989}
{Kirk} J.~G.,  {Heavens} A.~F.,  1989, \mnras, 239, 995

\bibitem[\protect\citeauthoryear{{Kuntz} \& {Snowden}}{{Kuntz} \&
  {Snowden}}{2000}]{KuntzSnowden2000}
{Kuntz} K.~D.,  {Snowden} S.~L.,  2000, \apj, 543, 195

\bibitem[\protect\citeauthoryear{{Kuntz} \& {Snowden}}{{Kuntz} \&
  {Snowden}}{2008}]{KuntzSnowden2008}
{Kuntz} K.~D.,  {Snowden} S.~L.,  2008, \aap, 478, 575

\bibitem[\protect\citeauthoryear{{Leccardi} \& {Molendi}}{{Leccardi} \&
  {Molendi}}{2008}]{LeccardiMolendi2008b}
{Leccardi} A.,  {Molendi} S.,  2008, \aap, 487, 461

\bibitem[\protect\citeauthoryear{{Macario}, {Markevitch}, {Giacintucci},
  {Brunetti}, {Venturi} \& {Murray}}{{Macario} et~al.}{2011}]{Macario2011}
{Macario} G.,  {Markevitch} M.,  {Giacintucci} S.,  {Brunetti} G.,  {Venturi}
  T.,    {Murray} S.~S.,  2011, \apj, 728, 82

\bibitem[\protect\citeauthoryear{{Markevitch}}{{Markevitch}}{2010}]{Markevitch%
2010}
{Markevitch} M.,  2010, ArXiv e-prints

\bibitem[\protect\citeauthoryear{{Markevitch} \& {Vikhlinin}}{{Markevitch} \&
  {Vikhlinin}}{2001}]{Markevitch2001}
{Markevitch} M.,  {Vikhlinin} A.,  2001, \apj, 563, 95

\bibitem[\protect\citeauthoryear{{Miniati}}{{Miniati}}{2002}]{Miniati2002}
{Miniati} F.,  2002, \mnras, 337, 199

\bibitem[\protect\citeauthoryear{{Moretti}, {Campana}, {Lazzati} \&
  {Tagliaferri}}{{Moretti} et~al.}{2003}]{Moretti2003}
{Moretti} A.,  {Campana} S.,  {Lazzati} D.,    {Tagliaferri} G.,  2003, \apj,
  588, 696

\bibitem[\protect\citeauthoryear{{Ogrean}, {Br{\"u}ggen}, {R{\"o}ttgering},
  {Simionescu}, {Croston}, {van Weeren} \& {Hoeft}}{{Ogrean}
  et~al.}{2012}]{Ogrean2012b}
{Ogrean} G.,  {Br{\"u}ggen} M.,  {R{\"o}ttgering} H.,  {Simionescu} A.,
  {Croston} J.,  {van Weeren} R.,    {Hoeft} M.,  2012, ArXiv e-prints

\bibitem[\protect\citeauthoryear{{Ogrean} \& {Br{\"u}ggen}}{{Ogrean} \&
  {Br{\"u}ggen}}{2012}]{Ogrean2012c}
{Ogrean} G.~A.,  {Br{\"u}ggen} M.,  2012, ArXiv e-prints

\bibitem[\protect\citeauthoryear{{Pinzke}, {Oh} \& {Pfrommer}}{{Pinzke}
  et~al.}{2013}]{Pinzke2013}
{Pinzke} A.,  {Oh} S.~P.,    {Pfrommer} C.,  2013, ArXiv e-prints

\bibitem[\protect\citeauthoryear{{R\"ottgering}, {Wieringa}, {Hunstead} \&
  {Ekers}}{{R\"ottgering} et~al.}{1997}]{Roettgering1997}
{R\"ottgering} H.~J.~A.,  {Wieringa} M.~H.,  {Hunstead} R.~W.,    {Ekers}
  R.~D.,  1997, \mnras, 290, 577

\bibitem[\protect\citeauthoryear{{Russell}, {Sanders}, {Fabian}, {Baum},
  {Donahue}, {Edge}, {McNamara} \& {O'Dea}}{{Russell}
  et~al.}{2010}]{Russell2010}
{Russell} H.~R.,  {Sanders} J.~S.,  {Fabian} A.~C.,  {Baum} S.~A.,  {Donahue}
  M.,  {Edge} A.~C.,  {McNamara} B.~R.,    {O'Dea} C.~P.,  2010, \mnras, 406,
  1721

\bibitem[\protect\citeauthoryear{{Ryu}, {Kang}, {Hallman} \& {Jones}}{{Ryu}
  et~al.}{2003}]{Ryu2003}
{Ryu} D.,  {Kang} H.,  {Hallman} E.,    {Jones} T.~W.,  2003, \apj, 593, 599

\bibitem[\protect\citeauthoryear{{Sanders} \& {Fabian}}{{Sanders} \&
  {Fabian}}{2012}]{Sanders2012}
{Sanders} J.~S.,  {Fabian} A.~C.,  2012, \mnras, 421, 726

\bibitem[\protect\citeauthoryear{{Sidher}, {Sumner}, {Quenby} \&
  {Gambhir}}{{Sidher} et~al.}{1996}]{Sidher1996}
{Sidher} S.~D.,  {Sumner} T.~J.,  {Quenby} J.~J.,    {Gambhir} M.,  1996, \aap,
  305, 308

\bibitem[\protect\citeauthoryear{{Simionescu}, {Allen}, {Mantz} \&
  {Werner}}{{Simionescu} et~al.}{2011}]{Simionescu2011}
{Simionescu} A.,  {Allen} S.,  {Mantz} A.,    {Werner} N.,  2011, in
  {J.-U.~Ness \& M.~Ehle} ed., The X-ray Universe 2011. p.~152

\bibitem[\protect\citeauthoryear{{Skillman}, {Xu}, {Hallman}, {O'Shea},
  {Burns}, {Li}, {Collins} \& {Norman}}{{Skillman} et~al.}{2013}]{Skillman2013}
{Skillman} S.~W.,  {Xu} H.,  {Hallman} E.~J.,  {O'Shea} B.~W.,  {Burns} J.~O.,
  {Li} H.,  {Collins} D.~C.,    {Norman} M.~L.,  2013, \apj, 765, 21

\bibitem[\protect\citeauthoryear{{van Weeren}, {Br{\"u}ggen}, {R{\"o}ttgering},
  {Hoeft}, {Nuza} \& {Intema}}{{van Weeren} et~al.}{2011}]{vanWeeren2011a}
{van Weeren} R.~J.,  {Br{\"u}ggen} M.,  {R{\"o}ttgering} H.~J.~A.,  {Hoeft} M.,
   {Nuza} S.~E.,    {Intema} H.~T.,  2011, \aap, 533, A35

\bibitem[\protect\citeauthoryear{{van Weeren}, {R{\"o}ttgering}, {Br{\"u}ggen}
  \& {Hoeft}}{{van Weeren} et~al.}{2010}]{vanWeeren2010}
{van Weeren} R.~J.,  {R{\"o}ttgering} H.~J.~A.,  {Br{\"u}ggen} M.,    {Hoeft}
  M.,  2010, Science, 330, 347

\bibitem[\protect\citeauthoryear{{van Weeren}, {R{\"o}ttgering}, {Intema},
  {Rudnick}, {Br{\"u}ggen}, {Hoeft} \& {Oonk}}{{van Weeren}
  et~al.}{2012}]{vanWeeren2012}
{van Weeren} R.~J.,  {R{\"o}ttgering} H.~J.~A.,  {Intema} H.~T.,  {Rudnick} L.,
   {Br{\"u}ggen} M.,  {Hoeft} M.,    {Oonk} J.~B.~R.,  2012, \aap, 546, A124

\bibitem[\protect\citeauthoryear{{Verner}, {Ferland}, {Korista} \&
  {Yakovlev}}{{Verner} et~al.}{1996}]{Verner1996}
{Verner} D.~A.,  {Ferland} G.~J.,  {Korista} K.~T.,    {Yakovlev} D.~G.,  1996,
  \apj, 465, 487

\end{thebibliography}

\label{lastpage}

\end{document}